\documentclass[aps,prb,twocolumn,superscriptaddress,notitlepage]{revtex4-1}
\usepackage{graphicx}
\usepackage[caption=false]{subfig}
\usepackage{float}
\usepackage{natbib}
\usepackage{xcolor}
\usepackage{xr}
\usepackage{amsmath}
\usepackage{amssymb}
\usepackage{array}
\usepackage{wasysym}
\usepackage{color,soul}
\usepackage{braket}
\usepackage{verbatim}
\usepackage{hyperref}
\usepackage{gensymb}
\usepackage{url}

\usepackage{filecontents}
\begin{document}

\title{Local nuclear and magnetic order in the two-dimensional spin glass, Mn$_{0.5}$Fe$_{0.5}$PS$_3$}

\author{J. N. Graham}
\affiliation{Department of Chemistry and Materials Innovation Factory, University of Liverpool, 51 Oxford Street, Liverpool, L7 3NY, UK}
\affiliation{Institut Laue-Langevin, $71$ Avenue des Martyrs, CS$20156$, $38042$ Grenoble C$\mathrm{\acute{e}}$dex $9$, France}
\affiliation{School of Chemistry, University of Birmingham, Edgbaston, Birmingham B15 2TT, UK}
\author{M. J. Coak}
\affiliation{Department of Physics, University of Warwick, Coventry, West Midlands, CV$4$ $7$AL, UK}
\author{S. Son}
\affiliation{Center for Correlated Electron Systems, Institute for Basic Science, Seoul $08826$, Republic of Korea}
\affiliation{Department of Physics and Astronomy, Seoul National University, Seoul $08826$, Republic of Korea}
\affiliation{Center for Quantum Materials, Seoul National University, Seoul $08826$, Republic of Korea}
\author{E. Suard}
\affiliation{Institut Laue-Langevin, $71$ Avenue des Martyrs, CS$20156$, $38042$ Grenoble C$\mathrm{\acute{e}}$dex $9$, France}
\author{J-G. Park}
\affiliation{Center for Correlated Electron Systems, Institute for Basic Science, Seoul $08826$, Republic of Korea}
\affiliation{Department of Physics and Astronomy, Seoul National University, Seoul $08826$, Republic of Korea}
\affiliation{Center for Quantum Materials, Seoul National University, Seoul $08826$, Republic of Korea}
\author{L. Clark}
\affiliation{Department of Chemistry and Materials Innovation Factory, University of Liverpool, 51 Oxford Street, Liverpool, L7 3NY, UK}
\affiliation{School of Chemistry, University of Birmingham, Edgbaston, Birmingham B15 2TT, UK}
\author{A. R. Wildes}
\affiliation{Institut Laue-Langevin, $71$ Avenue des Martyrs, CS$20156$, $38042$ Grenoble C$\mathrm{\acute{e}}$dex $9$, France}

\date{\today}

\begin{abstract}
\noindent
We present a comprehensive study of the short-ranged nuclear and magnetic order in the two-dimensional spin glass, Mn$_{0.5}$Fe$_{0.5}$PS$_3$. Nuclear neutron scattering data reveal a random distribution of Mn$^{2+}$ and Fe$^{2+}$ ions within the honeycomb layers, which gives rise to a spin glass state through inducing competition between neighbouring exchange interactions, indicated in magnetic susceptibility data by a cusp at the glass transition, $T_g = 35$~K. Analysis of magnetic diffuse neutron scattering data collected for both single crystal and polycrystalline samples gives further insight into the origin of the spin glass phase, with spin correlations revealing a mixture of satisfied and unsatisfied correlations between magnetic moments within the honeycomb planes, which can be explained by considering the magnetic structures of the parent compounds, MnPS$_3$ and FePS$_3$. We found that, on approaching $T_g$ from above, an ensemble-averaged correlation length of $\xi = 5.5(6)$~\r{A} developed between satisfied correlations, and below $T_g$, the glassy behaviour gave rise to a distance-independent correlation between unsatisfied moments. Correlations between the planes were found to be very weak, which mirrored our observations of rod-like structures parallel to the \textit{c*} axis in our single crystal diffraction measurements, confirming the two-dimensional nature of Mn$_{0.5}$Fe$_{0.5}$PS$_3$.
\end{abstract}
\maketitle

\section{Introduction}
\noindent
Low-dimensional materials, such as graphene, continue to captivate the scientific community due to the wide range of potential applications they naturally lend themselves to, from optoelectronics to nanocatalysis \cite{Tang2013,Wang2018,Ponraj_2016,Mak2016,Kumar2018}. Whilst their chemical and electronic properties have been studied extensively, examples of two-dimensional magnetic materials that can be exfoliated down to a monolayer, remain relatively scarce \cite{Burch2018,Park2016}. Yet lowering the dimensionality of certain compounds, and combining with, for example the application of pressure, has resulted in the formation of unconventional electronic and magnetic states of matter \cite{PhysRevLett.107.056802,Gong2017,Wang_2016,Huang2018,Coak2020,Haines2018} that are rarely encountered in their bulk counterparts. Another potential route to attain unconventional magnetic states is to study frustrated magnets \cite{Balents2010,Savary}. In recent years, there has been a surge of research focussed around novel ways to induce frustration on networks of magnetic ions that would not usually show such behaviour. One example is the Kitaev model \cite{Kitaev,kitaev2}, where highly anisotropic interactions between neighbouring spins generate frustration on a $S\mathrm{_{eff} } = 1/2$ honeycomb network. The search for the theoretically predicted Kitaev quantum spin liquid resulting from such frustrated interactions has so far centred around two main material families, $\alpha$ - RuCl$_3$ \cite{Banerjee2016,PhysRevB.91.241110,PhysRevB.91.144420,PhysRevB.93.075144} and $A_2$IrO$_3$ ($A$ = Li, Na) \cite{PhysRevB.93.195158,PhysRevB.85.180403,PhysRevB.83.220403}, where ultimately deviations from the ideal honeycomb and direct exchange between the transition-metal ions lead these materials to order. Alternatively, frustration may also be induced by increasing the strength of competing ferromagnetic and antiferromagnetic exchange interactions within the honeycomb network, such as in the Heisenberg antiferromagnet Bi$_3$Mn$_4$O$_{12}$(NO$_3$) \cite{PhysRevB.96.140404,PhysRevB.85.184412,PhysRevLett.105.187201,Okumura10}.
One family of materials that may link these fields are the honeycomb layered metal thiophosphates, \textit{M}PS$_3$.\\
\begin{figure}
\includegraphics[width = 0.5\textwidth]{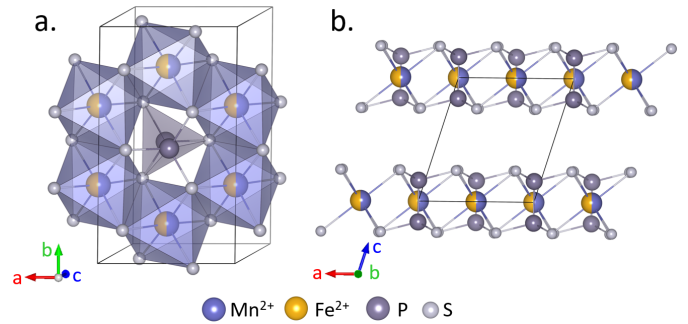}
\caption{Mn$_{0.5}$Fe$_{0.5}$PS$_3$ adopts a $C2/m$ crystal structure where \textbf{a.} honeycomb layers of transition metal ions form in the \textit{ab} planes and are \textbf{b.} weakly bound by van der Waals interactions along the $c$ axis. Figures produced in VESTA \cite{Vesta}.}
\label{PS3structure}
\end{figure}
\\
The \textit{M}PS$_3$ compounds adopt a monoclinic $C2/m$ structure in which honeycomb networks of transition-metal (\textit{M}) ions form in the \textit{ab} planes\cite{Ouvrard1985,Brec1979} (Figure. \ref{PS3structure}a.). Each transition-metal ion is contained within an octahedron of sulphur atoms with a pair of phosphorous atoms found at the centre of each honeycomb ring, stacked in an ABC sequence, similar to the CdCl$_2$ structure \cite{Wang2018}. These layers are weakly bound by van der Waals interactions \cite{Ouvrard1985} (Figure. \ref{PS3structure}b.), allowing the layers to either be intercalated \cite{Brec1979,Foot1987,Grasso} or delaminated to a single monolayer \cite{Kuo2016,Lee2016}.\\
\\
The magnetic structures of two members of this series, MnPS$_3$ and FePS$_3$, have been well characterised in a number of earlier studies \cite{Wang2018,PhysRevB.46.5134,Susner2017,Brec1979,Grasso}. Although both are quasi-two-dimensional antiferromagnets, the precise details of their respective magnetic structures are deeply reliant on the choice of transition-metal ion residing within the honeycomb layers. Notably, in MnPS$_3$, all nearest-neighbour magnetic moments are coupled antiferromagnetically in the \textit{ab} planes, whereas in FePS$_3$, there are two ferromagnetic nearest-neighbours and one antiferromagnetic neighbour for every Fe$^{2+}$ ion within the plane \cite{PhysRevB.82.100408,Lancon2016,Murayama2016,Wildes_1998,Kurosawa83}. Furthermore, MnPS$_3$ is well modelled by a Heisenberg Hamiltonian, whilst the magnetic moments in FePS$_3$ are more appropriately described as Ising-like \cite{PhysRevB.46.5134,Lancon2016,Wildes_1998}.\\
\\
Creating a solid solution of these compounds, therefore, presents another way to generate magnetic frustration on a honeycomb network. Here, the source of frustration originates from the first nearest-neighbour exchange interaction, which is antiferromagnetic for MnPS$_3$ \cite{Wildes_1998} and ferromagnetic for FePS$_3$ \cite{Lancon2016}. This frustration may be further compounded by the competition between spin and exchange anisotropies that exist within the magnetic structures of the two end-members. Previous magnetisation and specific heat capacity measurements of Mn$_{0.5}$Fe$_{0.5}$PS$_3$ have revealed a spin glass phase exists with the glass transition, $T_g$, occurring at $35$~K \cite{Masubuchi2008,Manriquez2000,He2003,Takano2003}. This glassy phase, where frustrated moments are randomly frozen, is thought to arise as a result of the competing magnetic exchange interactions combined with the random site occupancy of the transition-metal ions within the honeycomb layers. The presence of such chemical disorder has been verified using techniques such as M$\mathrm{\ddot{o}}$ssbauer spectroscopy \cite{He2003}. Other sister compounds have also been reported, for instance Mn$_{0.5}$Fe$_{0.5}$PSe$_3$ \cite{Bhutani2020}, which despite the presence of chemical disorder and similarity to our material, is not a spin glass due to the very strong spin anisotropy arising from ligand spin-orbit contributions. This results in short-range magnetic order existing between nanoclusters of MnPSe$_3$ and FePSe$_3$-type structures.\\
\\
However, in the case of Mn$_{0.5}$Fe$_{0.5}$PS$_3$, there are still a number of important details surrounding the spin glass phase that remain to be understood, in particular, the nature of its local nuclear and magnetic correlations. In this paper we address these issues through a series of neutron scattering and magnetisation measurements.

\section{Experimental Methods }
\noindent
Single crystals of Mn$_{0.5}$Fe$_{0.5}$PS$_3$ were grown via a three-step process to maximise the homogeneous mixing of manganese and iron constituents. Stoichiometric quantities of elemental powders - Fe (99.998~\%, Alfa Aesar), Mn (99.95~\%, Alfa Aesar), P (red phosphorous, 99.99~\%, Sigma Aldrich) and S (99.998~\%, Alfa Aesar) were ground under an argon atmosphere until fine and sealed in quartz tubes, which were evacuated to a $10$~mbar Ar gas pressure. Quartz tubes used in each step had an $18$~mm inner diameter and were $100$~mm in length. The sealed tubes were annealed at $500$~$\mathrm{^oC}$ for $2$~days. The powder from this initial heating stage was then reground in an Ar glove box. The ground powder was sealed in a quartz tube as per the first synthesis step, and annealed at $600$~$\mathrm{^oC}$ for a further $2$ days. Single crystals were grown from the powder via chemical vapour transport. The powder was reground and loaded into quartz tubes with an iodine powder flux agent ($50$~mg of I$_2$ per $1$~g of reactants). The tubes were evacuated to $\sim 10^{-3}$~mbar vacuum using a molecular diffusion pump and sealed. The sealed tubes were placed in a two-zone furnace and heated to $730$ $\mathrm{^oC}$/$630$~$\mathrm{^oC}$ over $12$~hours and held for $7$~days before cooling to room temperature over $24$~hours. The grown crystals formed shiny black flakes with typical dimensions, $1$ cm $\times$ $1$ cm $\times$ $50$~$\mu$m. The sample stoichiometry and quality were verified with energy dispersive X-ray (EDX) spectroscopy and powder X-ray diffraction (XRD). EDX was performed on a Bruker QUANTAX system combined with a scanning electron microscope (COXEM, EM-30). The quality of single crystals were verified via X-ray Laue diffraction, using an Imaging Plate X-Ray Diffraction (IP-XRD) Laue Camera (IPX Co. Ltd).\\
\\
Constant wavelength neutron powder diffraction (NPD) data were collected at room temperature on the high-resolution D$2$B diffractometer \cite{D2Bpaper} at the Institut Laue-Langevin (ILL), Grenoble \cite{D2Bdata}. The incident wavelength was $\lambda = 1.59432$~\r{A} and the scattering was measured over an angular range of $10 < 2\theta < 160$~$\mathrm{^o}$. Nuclear refinements were completed using the GSAS package \cite{GSAS2}. The powder used in these measurements was the powder produced from step two of the growth process. Like previous members of the \textit{M}PS$_3$ family \cite{Wildes2015}, preferred orientation had to be accounted for due to the plate-like nature of Mn$_{0.5}$Fe$_{0.5}$PS$_3$ which makes grinding an isotropic powder difficult, and was corrected for via the spherical harmonics (ODF) method.\\
\\
Magnetisation measurements were carried out on single crystal samples in MPMS$3$ and MPMS SQUID magnetometers, Quantum Design. DC measurements were carried out in a fixed-$0.1$ T field, both cooled in zero applied field (ZFC) and field cooled (FC) cycles over a temperature range of $2$ to $300$~K. To avoid causing strains, the sample was not held in any epoxy or grease, and was instead held between sheets of plastic film. \\
\\
Single crystals were aligned in orientations parallel and perpendicular to the \textit{c*} axis on the three-axis spectrometer IN$3$ at the ILL. Neutron scattering measurements were performed on powder and single crystal samples using the polarised diffuse scattering D$7$ diffractometer \cite{Stewart:db5048} at the ILL \cite{D7data}. The incident neutron wavelength was $\lambda = 4.8718$~\r{A}, giving a reciprocal space range of $0.15 \leq Q \leq 2.5$ ~\r{A}$^{-1}$. Single crystal data were measured as a function of the sample rotation about the normal to the scattering plane. The samples were rotated over a total range of $210$ degrees. The technique of $xyz$-polarisation analysis was used to separate the magnetic, nuclear-coherent (NC) and nuclear spin-incoherent (NSI) components \cite{Stewart:db5048,Schweika_2010}. Full calibration measurements included measurements of an empty and cadmium-filled sample can to estimate the instrumental background, a quartz standard to correct for polarisation inefficiencies and a vanadium standard to normalise the detector efficiencies. Data were placed on an absolute intensity scale (with units b/ster/f.u.) by normalising to    the incoherent scattering from the vanadium standard. We verified our normalisation process was correct through Rietveld analysis of the NC cross-sections using FullProf \cite{Fullprof}. Magnetic diffuse scattering data were analysed using the programs SPINVERT \cite{PhysRevLett.108.017204,Paddison_2013}, SPINCORREL and Scatty \cite{PaddisonScatty}. These software produce refinements to experimental data based on reverse Monte Carlo (RMC) spin simulations. SPINVERT calculations were repeated $10$ times to average out statistical noise.

\section{Results}
\subsection{Nuclear Structure}
\begin{figure}
\centering
\includegraphics{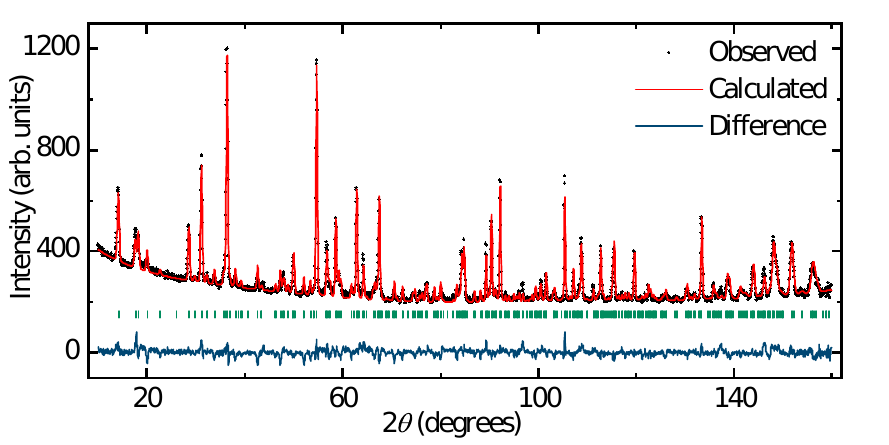}
\caption{Rietveld refinement of the $C2/m$ model ($\chi^2=5.273$, $\mathrm{R_{wp}}=4.54\%$) describing the structure of Mn$_{0.5}$Fe$_{0.5}$PS$_3$ using data collected on the D$2$B instrument at $300$ K. }
\label{D2Bfigure}
\end{figure}

\begin{figure}
\centering
\includegraphics{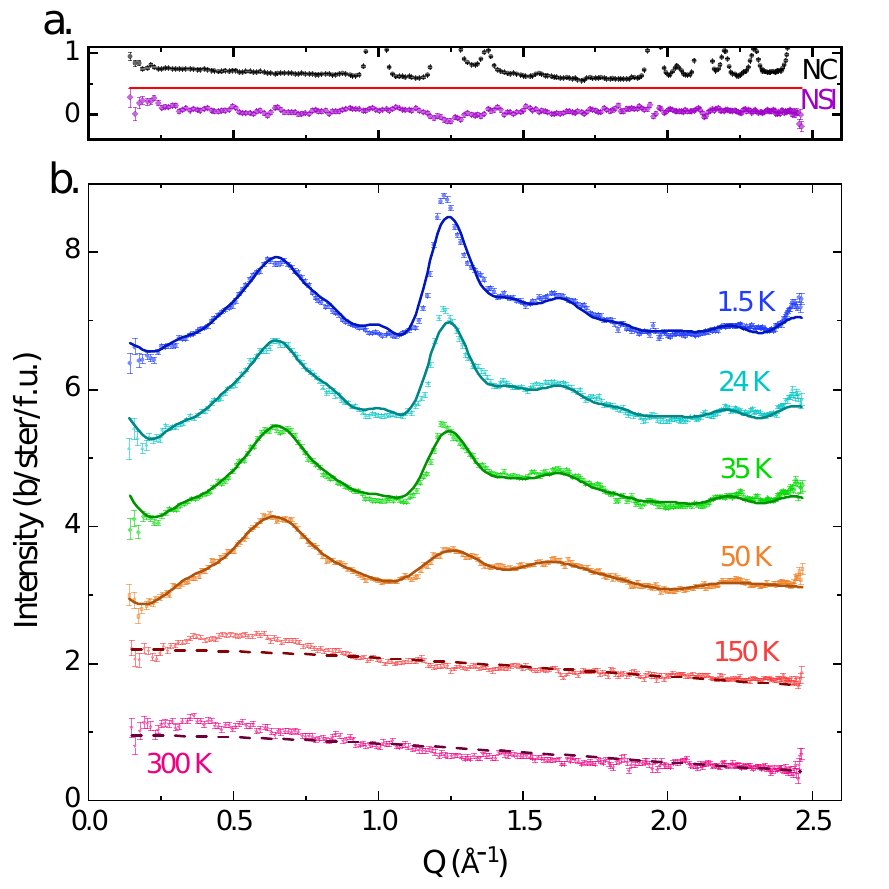}
\caption{Experimental powder neutron scattering data from D$7$. \textbf{a.} NC and NSI components measured at $1.5$ K. The expected NC is highlighted by the red line and is slightly lower than the measured value. \textbf{b.} Magnetic diffuse scattering contributions. These data have been vertically shifted by $1.25$ b/ster/f.u. for clarity. Fits produced by a RMC refinement assuming moments lie along the \textit{c*} axis are shown by the solid lines and fits considering just the paramagnetic contribution of Mn$^{2+}$ ions are shown by the dashed lines.}
\label{Ising}
\end{figure}

\noindent
Figure. \ref{D2Bfigure} shows the Rietveld refinement of the monoclinic $C2/m$ model against high-resolution NPD data of Mn$_{0.5}$Fe$_{0.5}$PS$_3$ collected at $300$ K on the D$2$B instrument at the ILL. The refined parameters can be found in Table \ref{D2BTable}. Refinement of the transition-metal occupancies confirmed that the sample has the expected stoichiometry whereby the honeycomb network is comprised of an approximately equal amount of Mn$^{2+}$ and Fe$^{2+}$ ions. Within error, this is in agreement with the EDX results of $47(1) \%$ and $53(2) \%$ for Mn$^{2+}$ and Fe$^{2+}$ respectively. The respective neutron scattering lengths for Mn$^{2+}$ and Fe$^{2+}$ are $b_{\mathrm{Mn}} = -3.73$~fm and $b_{\mathrm{Fe}} = +9.45$~fm, which yield excellent contrast. The absence of any diffuse structure within the flat background of the NPD data indicates a lack of any short-range ordering between the Mn$^{2+}$ and Fe$^{2+}$ ions, and therefore implies a random distribution of the magnetic ions throughout the average structure.\\
\begin{table}
\centering
\begin{tabular} { c  c  c  c  c  c c}
\hline
\hline
Atom & Site & x & y & z & Occ & U$\mathrm{_{iso}} $ (\r{A}$^2$)\\
\hline
\hline
Mn1 & 4\textit{g} & 0 & 0.3354(9) & 0 & 0.520(4) & $0.022(2)$\\
Fe1 & 4\textit{g} & 0 & 0.3354(9) & 0 & 0.480(4)& $0.022(2)$\\
P1 & 4\textit{i} & 0.0542(8) & 0 & 0.1619(5) & 1.0 & $0.0188(7)$\\
S1 & 4\textit{i} & 0.738(1) & 0 & 0.2442(8) & 1.0 & $0.0051(5)$\\
S2 & 8\textit{j} & 0.2358(7) & 0.1661(4) & 0.2503(6) & 1.0 & $0.0051(5)$\\
\hline
\hline
\end{tabular}
\caption{Refined nuclear structure parameters for the $C2/m$ model fitted to NPD data collected at
300 K, where the refined lattice parameters are \textit{a} = 6.0137(2) \r{A}, \textit{b} $ = 10.4174(2)$ \r{A} and \textit{c} = 6.7591(2) \r{A}, $\beta = 107.129(2)$ $ \mathrm{^o}$.}
\label{D2BTable}
\end{table}
\\
Experimental powder neutron scattering data recorded on D$7$ at the ILL are presented in Figure \ref{Ising}. An example of the NC and NSI components are shown in Figure \ref{Ising}a. which were collected at $1.5$ K. The expected NC is slightly lower than the measured values, as shown by the red line, and this may be due to a small systematic error in the estimation of the background. This is possible when using the \textit{xyz}-method, although it is noteworthy that any residual background will only appear in the NC and NSI contributions as it is self-subtracted from the magnetic contributions. Similarly to the D$2$B data, there is a nearly flat background in the NC measurement, which further confirms the absence of any substantial short-ranged order and therefore an essentially random distribution of Mn$^{2+}$ and Fe$^{2+}$ ions exists within the honeycomb layers.
\subsection{Magnetic Susceptibility }
\begin{figure}
\centering
\includegraphics{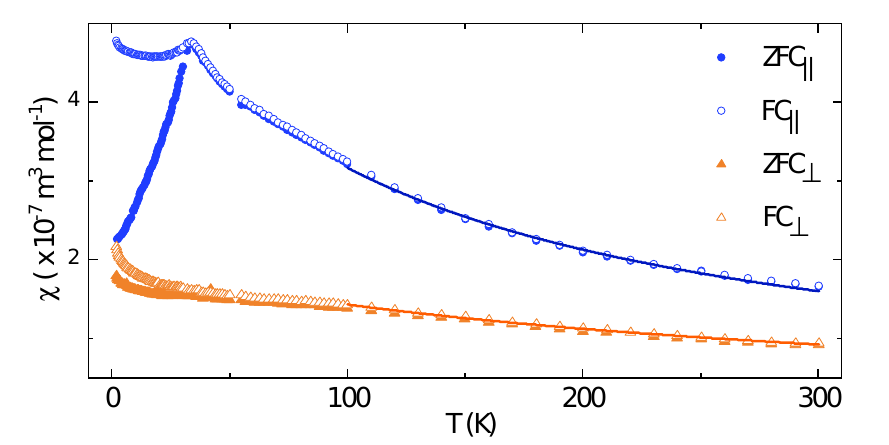}
\caption{DC Susceptibility data for Mn$_{0.5}$Fe$_{0.5}$PS$_3$ in a $0.1$ T magnetic field. Curie-Weiss fits were conducted over a range of $100$ to $300$ K and are shown by the blue and orange lines for when the \textit{c*} axis is parallel and perpendicular to the applied field respectively.}
\label{susc}
\end{figure}

\noindent
Single crystal magnetisation data are presented in Figure \ref{susc} and reveal features concomitant with spin glass behaviour. First, the data show a sharp cusp at $35$~K, which corresponds to the glass transition, $T_g$, and is in accordance with other reports \cite{Masubuchi2008,Takano2003}. Splitting between the FC and ZFC magnetisation occurs below $T_g$, which is particularly prominent when the applied field is parallel to the \textit{c*} axis. These data have been fit between $100$ and $300$~K with a modified Curie-Weiss law, $\chi = \frac{C}{T-\theta} + \chi_{0}$, where $\chi_{0}=-1.0631\times10^{-9}$ ~m$^3$~mol$^{-1}$ is a temperature independent background term describing the inherent diamagnetic contribution of Mn$_{0.5}$Fe$_{0.5}$PS$_3$\cite{Bain2008}. When the \textit{c*} axis is parallel and perpendicular to the applied field, the extracted Curie-Weiss constants are $\theta_{||} = -105(4)$~K and $\theta_{\perp} = -267(4)$~K respectively. These values are both large and negative, indicating that the dominant interactions in Mn$_{0.5}$Fe$_{0.5}$PS$_3$ are antiferromagnetic. The Curie constants, $C_{||} = 6.5(1) \times 10^{-5}$~m$^{3}$ K mol$^{-1}$ and $C_\perp = 5.29(5)\times10^{-5}$~m$^{3}$ K mol$^{-1}$, give effective magnetic moments, $\mathrm{\mu_{eff} =6.43(1)} \mathrm{\mu_B}$ and $\mathrm{\mu_{eff} = 5.80(3)} \mathrm{\mu_B}$ when the applied field is parallel and perpendicular to the \textit{c*} axis respectively. This is slightly larger than the expected spin-only effective moment, $\mathrm{\mu_{eff} = 5.41} \mathrm{\mu_B}$, assuming an average $S = 2.25$ from the mixture of Mn$^{2+}$ and Fe$^{2+}$ ions. Although previous magnetisation studies of MnPS$_3$ have shown that the system can be well modelled by a spin-only effective moment, the high-spin state of Fe$^{2+}$ ions results in a sizeable spin-orbit contribution, seen in both magnetisation \cite{PhysRevB.46.5134} and neutron \cite{Lancon2016} studies of FePS$_3$. The large difference in the measured susceptibility when the field is applied in different orientations indicates that Mn$_{0.5}$Fe$_{0.5}$PS$_3$ is highly anisotropic, similar to the measured magnetic susceptibility of FePS$_3$ with pronounced Ising anisotropy parallel to the \textit{c*} axis.
\subsection{Magnetic Diffuse Scattering }
\noindent
The magnetic cross-sections recorded on D$7$ are presented in Figure \ref{Ising}b. and have been vertically shifted by $1.25$ b/ster/f.u. for clarity. The high-temperature data ($\geq 150$ K) have been fit using an analytical approximation of the magnetic form factor for Mn$^{2+}$ ions\cite{Magff}, as shown by the dashed lines. This function describes the expected nature of Mn$^{2+}$ ions when in a purely paramagnetic state. In the paramagnetic regime, the analytical approximations of Mn$^{2+}$ and Fe$^{2+}$ are roughly equivalent, therefore to simplify the analysis only one magnetic ion has been used. This function compares with the data reasonably well, however, the associated paramagnetic moment at $Q = 0$~\r{A}$^{-1}$ is $\mu_\mathrm{eff} = 4.45 \mu_\mathrm{B}$ which is a little low when compared to the expected $\mu_{\mathrm{eff}} = 5.41 \mu_\mathrm{B}$. This can occur when there is some additional inelastic scattering at energies outside the integration window of D$7$, which may be attributed to persistent critical fluctuations that extend beyond $20$~meV, similar to those found in FePS$_3$ \cite{Wildes12}. The noticeable deviation between the function and data, particularly at low $Q$, indicates that some residual short-range correlations are still present, even at $300$~K.\\
\\
Below $50$ K, there are a mixture of broad and sharp features which are reminiscent of the parent compounds, MnPS$_3$ and FePS$_3$. For instance, the broad feature centred around $Q = 0.6$~\r{A}$^{-1}$ is similar to that observed in FePS$_3$ \cite{Lancon2016,Goossens2000,Rule2002,Rule2003}. The magnitude of this feature has little variation below $50$ K and therefore suggests that FePS$_3$-type correlations are more or less fully developed above $T_g$. Furthermore, the sharp peak at $Q = 1.2$~\r{A}$^{-1}$ is similar to that observed in MnPS$_3$ \cite{Wildes1994,Rule2003}. This feature continues to evolve with temperature, thus implying that MnPS$_3$-type correlations are still developing below $T_g$. These data were fit using the SPINVERT program. A box size of $12 \times 12 \times 12$ unit cells were constructed with the moments constrained to lie either parallel or antiparallel to the \textit{c*} axis. The resultant reverse Monte Carlo (RMC) fits are shown in Figure \ref{Ising}b. by the solid lines. The assumption that moments lie along the \textit{c*} axis is supported by the anisotropy in the paramagnetic susceptibility and the fact that the orientation of ordered moments in the parent compounds align approximately along the \textit{c*} axis. Specifically in MnPS$_3$ the moments are canted $7$ $\mathrm{^o}$ from the \textit{c*} axis, whereas in FePS$_3$ ordered moments lie directly along the \textit{c*} axis  \cite{Susner2017,PhysRevB.82.100408}. The SPINVERT fits compare to the powdered data well, as they maintain a good balance between fitting all the features, both broad and sharp, within the data. The simplicity of this model is particularly appealing, by allowing the moments to lie only collinear to the \textit{c*} axis, we were able to not only model these data well but also extend this to our single crystal study. Modelling with more complex models, with more free parameters such as larger box sizes or more rotational degrees of freedom for the magnetic moments, did not produce as good results as the ones presented here, as these additional parameters began to fit the statistical noise. We notice some small anomalies to the fit, such as the small peak emerging at $Q = 0.9$~\r{A}$^{-1}$ and that the intensity of the sharp peak at $Q = 1.2$~\r{A}$^{-1}$ is lower than the measured value, however we are confident in this model as the extracted magnetic moment from SPINVERT, $\mu_\mathrm{{eff}} = 5.23(8)\mu_\mathrm{B}$ ($S = 2.15(4)$) is in good agreement with the expected moment, $\mu_{\mathrm{eff}} = 5.41 \mu_\mathrm{B}$ ($S = 2.25$). Additionally, the quality of our later reconstruction of single crystal data indicates that this model is a reliable estimate of the magnetic diffuse scattering of Mn$_{0.5}$Fe$_{0.5}$PS$_3$.\\
\begin{figure}
\includegraphics{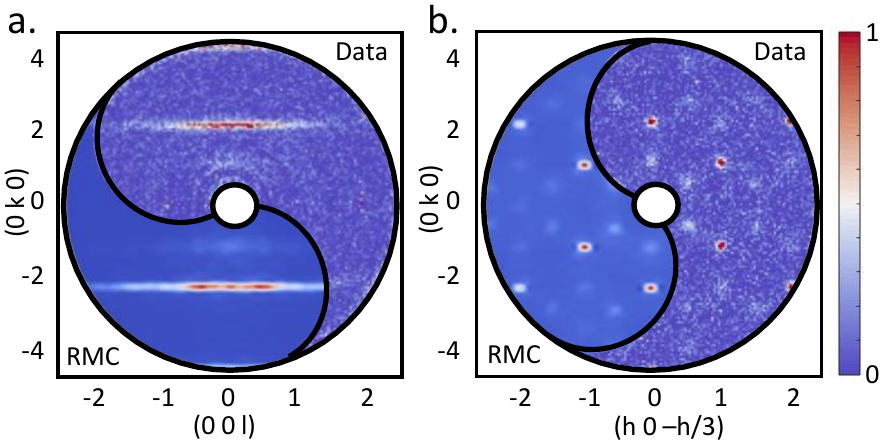}
\caption{Magnetic component of single crystal neutron scattering data measured at $1.5$ K compared against the RMC SPINVERT model. \textbf{a.} Data in the (0 0 l), (0 k 0) plane reveal rod-like structures that reflect the two-dimensionality of Mn$_{0.5}$Fe$_{0.5}$PS$_3$ and \textbf{b.} data in the (h 0 $\frac{-h}{3}$) plane, which is orthogonal to the plane in part \textbf{a.}, shows strong magnetic intensity at nuclear Bragg peak positions. The intensity of these data have been normalised to the maximum of the measured intensity as given by the colour bar.}
\label{SC}
\end{figure}
\\
Figure \ref{SC} shows the measured single crystal data from D$7$ at $1.5$ K. Figure \ref{SC}a., shows rod-like scattering parallel to the \textit{c*} axis. These observations coincide with the main feature in our low-temperature powder data, the sharp peak at $Q = 1.2$ \r{A}$^{-1}$. Both of these features are synonymous with two-dimensional structures and thus reflect the two-dimensional nature of Mn$_{0.5}$Fe$_{0.5}$PS$_3$. When the \textit{c*} axis is normal to the scattering plane, as in Figure \ref{SC}b., we observe strong magnetic intensity, approximately $60$ degrees apart, which correspond to the expected magnetic Bragg positions for ordered MnPS$_3$ \cite{PhysRevB.82.100408}. Additional magnetic Bragg spots are observed at low-temperatures, with the same six-fold rotation, but their intensity is weaker. Single crystal diffraction patterns were produced from fits of our powder data by the Scatty program. A comparison between the experimental data and scattering predicted from the RMC fits is shown in Figure \ref{SC}, the good agreement indicates that the fit results will give a reliable estimate of the spin correlations in Mn$_{0.5}$Fe$_{0.5}$PS$_3$.

\subsection{Spin Correlations }

\begin{figure}
\includegraphics{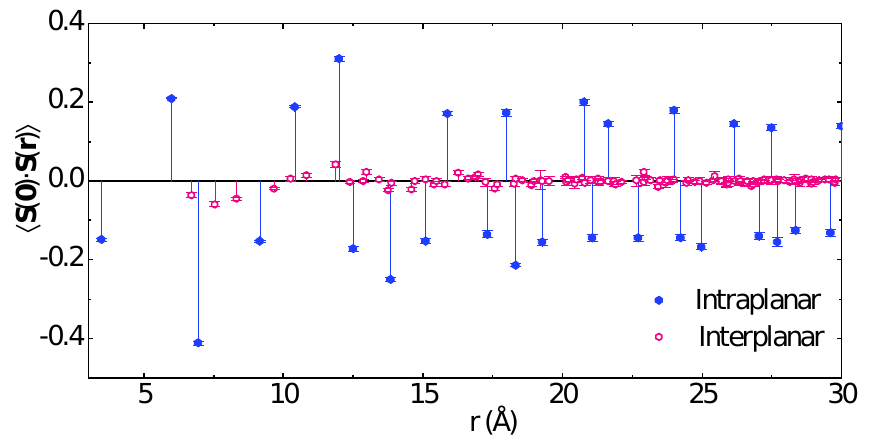}
\caption{Normalised spin-spin correlation function $\langle \textbf{S}(\textbf{0}) \cdot \textbf{S}(\textbf{r}) \rangle$ values for increasing interatomic distances at $1.5$ K. Correlations have been separated into intra- and interplanar by blue and pink markers respectively. }
\label{Correlations1}
\end{figure}
\begin{figure*}
\includegraphics[width = \textwidth]{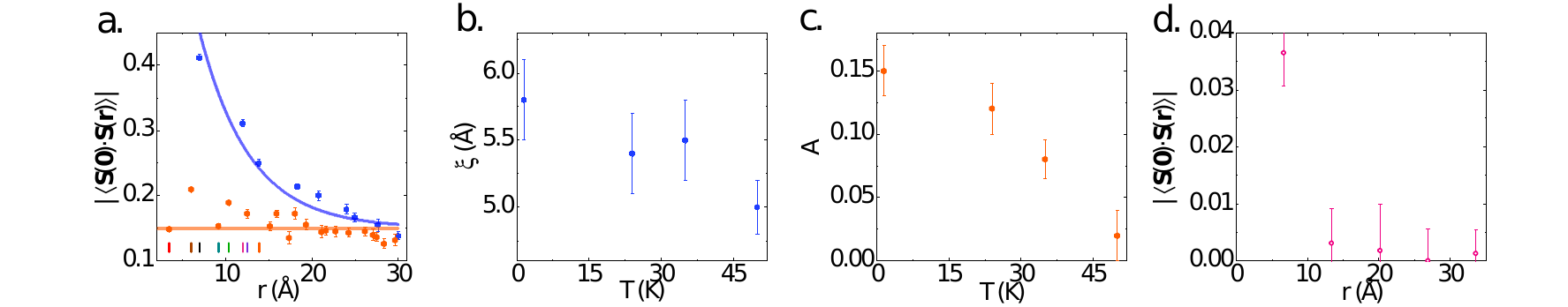}
\caption{\textbf{a.} Magnitude of intraplanar correlations, $|\langle\textbf{S(0)}\cdot\textbf{S(r)}\rangle|$ at $1.5$ K. Blue markers highlight correlations that decrease in magnitude with increasing \textbf{r} and orange markers highlight correlations that remain roughly constant in magnitude with increasing separation. These data have been fit with an exponential and linear function as shown by the blue and orange lines respectively. The first eight correlations have an associated coloured tickmark which links to a corresponding arrow in Figure~\ref{Correlations3}. \textbf{b.} Extracted correlation length, $\xi$, from the exponential fit as a function of temperature is roughly constant below $T_g$. \textbf{c.}~Linear fits, $|\langle\textbf{S(0)} \cdot\textbf{S(r)}\rangle| = A$, as a function of temperature decrease in magnitude with increasing temperature, becoming zero above $T_g$. \textbf{d.} Magnitude of interplanar correlations, $|\langle\textbf{S(0)}\cdot\textbf{S(r)}\rangle|$, lying closest to the \textit{c*} axis at $1.5$ K. Measurable intensity is only observed between the first set of adjacent planes.}
\label{Correlations2}
\end{figure*}
\noindent
The radial spin correlation functions, $\langle \textbf{S(0)} \cdot \textbf{S(r)} \rangle$, were calculated by SPINCORREL, where $\langle \textbf{S(0)} \cdot \textbf{S(r)} \rangle$ is the scalar product of a normalised spin with its neighbour at a vector separation, $\textbf{r}$ \cite{Paddison_2013}. The function $\langle \textbf{S(0)} \cdot \textbf{S(r)} \rangle$ is normalised to equal $+1$ if all the neighbours at a distance \textbf{r} are ferromagnetically coupled, and equal to $-1$ if all the neighbours are antiferromagnetically coupled. Figure \ref{Correlations1} shows $\langle \textbf{S(0)} \cdot \textbf{S(r)} \rangle$ at $1.5$ K split between intra- and interplanar correlations by blue and pink markers respectively.\\
\\
Figure \ref{Correlations2}a. shows the magnitude, $|\langle \textbf{S(0)} \cdot \textbf{S(r)} \rangle|$, of the intraplanar correlations where two trends are apparent. The first set of correlations, highlighted by the orange markers, are approximately constant as a function of interatomic spacing, $\textbf{r}$. The second, shown by the blue markers, decrease in magnitude with increasing $\textbf{r}$. These two distinct trends in the data were analysed using two different functions. The orange points were averaged out to give a constant, $|\langle \textbf{S(0)} \cdot \textbf{S(r)} \rangle|=A$, and the blue markers were fit with an exponential function, $|\langle \textbf{S(0)} \cdot \textbf{S(r)} \rangle| = \mathrm{exp}\large(-\frac{\textbf{r}}{\xi}\large) + A$, where $\xi$ is the characteristic spin-spin correlation length (\r{A}). The fits for the $1.5$ K data are shown in Figure \ref{Correlations2}a. Intraplanar correlations were found to be temperature dependent, and the fit values of $\xi$ and $A$ as a function of temperature are shown in Figures \ref{Correlations2}b. and c. respectively.\\
\\
Figure \ref{Correlations2}d. shows the magnitude of the interplanar correlations for those neighbours closest to the \textit{c*} axis, that is the nearest neighbours between parallel planes. The magnitude of these correlations showed no noticeable change with temperature. These correlations are very weak, only the first nearest-neighbour, at $\textbf{r}=6.71$~\r{A}, has a magnitude significantly greater than zero and that is small at $|\langle\textbf{S(0)}\cdot\textbf{S(r)}\rangle| = 0.035$. This weak interaction between adjacent planes reflects the two-dimensional nature of Mn$_{0.5}$Fe$_{0.5}$PS$_3$.

\section{Discussion}

\begin{figure}
\includegraphics[width = 0.5\textwidth]{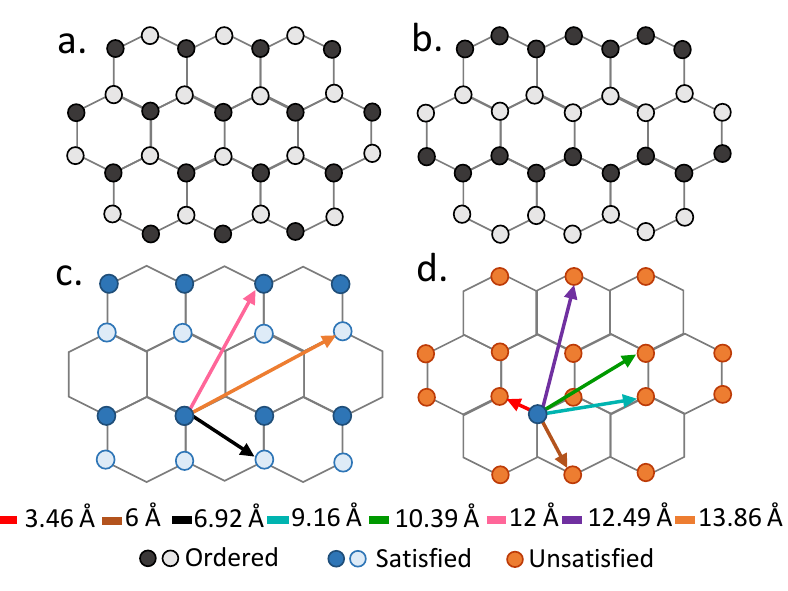}
\caption{In-plane magnetic structures of \textbf{a.} MnPS$_3$ and \textbf{b.} FePS$_3$ have conventional long-ranged ordered states. Spin up and spin down moments in these antiferromagnetic structures are indicated by closed and open circles respectively. \textbf{c.} Common orientations of magnetic moments between MnPS$_3$ and FePS$_3$ are present in the mixed Mn$_{0.5}$Fe$_{0.5}$PS$_3$, and the correlations between these moments have be considered as satisfied. \textbf{d.} All other moment orientations are competing between the two structures, and so the correlations between these moments have been regarded as unsatisfied. The first eight distances have been plotted from an arbitrary point, where the colour of the arrow links to a corresponding tickmark in Figure \ref{Correlations2}a.}
\label{Correlations3}
\end{figure}
\noindent
One plausible explanation as to why these two trends exist in Figure \ref{Correlations2}a. can be understood by considering the magnetic structures of MnPS$_3$ and FePS$_3$. In MnPS$_3$, all the nearest-neighbour interactions within the \textit{ab} planes are antiferromagnetic, whereas in FePS$_3$, there are two ferromagnetic neighbours and one antiferromagnetic neighbour. This is illustrated in Figure \ref{Correlations3}a. and b. for MnPS$_3$ and FePS$_3$ respectively, where closed circles represent a spin-up orientation and open circles represent a spin-down orientation of the magnetic moments within the magnetic structures. Despite the differences in the magnetic structures, there are certain magnetic moment orientations that are the same, for which the underlying correlations must be similar. These correlations can thus be considered as satisfied and are illustrated in Figure \ref{Correlations3}c. Each satisfied correlation gives rise to a defined spin orientation between pairs of magnetic moments, just like in a conventionally ordered magnetic ground state, as shown by the open and closed blue circles. The remaining correlations are competing, and therefore unsatisfied, which are represented by the orange circles in Figure \ref{Correlations3}d. Figures \ref{Correlations3}c. and d. show the first eight correlations as given by SPINCORREL plotted from an arbitrary site, where it can be seen that these correlations fall into the two aforementioned groups. Some correlations could fit into either group, such as the shortest interatomic distance, $\textbf{r} = 3.46$~\r{A}, which has three neighbours. Of the three, one correlation is satisfied and the other two are unsatisfied. All neighbours at the distances marked in Figure \ref{Correlations3}c. have solely satisfied correlations. If any neighbour in a group of equivalent distances has an unsatisfied interaction then it has been assigned to the unsatisfied group in Figure \ref{Correlations3}d., regardless of how many satisfied correlations that distance may have. Therefore, the correlation occurring at $\textbf{r} = 3.46$~\r{A} has been assigned to the unsatisfied group. These groups are in perfect agreement when compared back to the observed trends in Figure \ref{Correlations2}a. To illustrate this, we have related the colour of tickmark for the first eight correlations in Figure \ref{Correlations2}a. with its corresponding arrow in Figure \ref{Correlations3}.\\
\\
The average correlation length between satisfied moments is approximately constant below the freezing temperature as shown in Figure \ref{Correlations2}b. The average correlation length of these correlations, $\xi = 5.5(6)$~\r{A} is relatively small, as it does not even stretch one unit cell across in the honeycomb. So although there is some short-ranged magnetic order present, ordered interactions within the plane can only be considered between the very closest neighbouring ions, and long-ranged magnetic order does not develop across the honeycomb network. In Figure \ref{Correlations2}c., we observe that the dropoff of $A$ is more pronounced with increasing temperature, becoming approximately zero just above $T_g$. We can therefore relate that the unsatisfied moments give rise to the glassy behaviour of Mn$_{0.5}$Fe$_{0.5}$PS$_3$ and observe that they reach a local level of ordering with continued cooling.\\
\\
To summarise, the correlations can be separated into three groups.  The first, interplanar correlations, are very weak, verifying the two-dimensional character of the compound.  The intraplanar correlations may be separated into two subgroups: satisified correlations, between moments that have the same orientations for all equivalent distances in both the MnPS$_3$ and FePS$_3$ structures; and unsatisfied correlations, where moments at equivalent distances have at least one pair that differ between the two structures.  The unsatisfied correlations are approximately constant with distance and are temperature-dependent, falling to zero above $T_g$.  The satisfied correlations decrease exponentially with distance, having a characteristic length that is much less temperature-dependent and that persists above the glass temperature.

\section{Conclusion}
\noindent
In conclusion, we have presented high resolution powder neutron diffraction, DC magnetic susceptibility and magnetic diffuse scattering data for the two-dimensional honeycomb spin glass, Mn$_{0.5}$Fe$_{0.5}$PS$_3$. We have confirmed that no short- or long-ranged nuclear ordering is present within the honeycomb layers from the random distribution of Mn$^{2+}$ and Fe$^{2+}$ ions. Neutrons were the ideal tool for determining this, as the difference in neutron scattering lengths of Mn$^{2+}$ and Fe$^{2+}$ provided a clear contrast, which is not resolvable in other methods, like X-ray diffraction. The presence of the spin glass phase has been validated through DC magnetic susceptibility measurements, through the observation of a characteristic sharp cusp at the glass transition. In modelling and interpreting our magnetic diffuse scattering data we have relied upon the nature of the magnetic structure of the parent compounds, MnPS$_3$ and FePS$_3$. Whilst the differences produced unsatisfied interactions that cause glassy behaviour, similarities were found to be responsible for some local magnetic ordering, both of which are required to understand two emergent trends in the data collected. We confirmed this model through the excellent agreement with measured single crystal data. We have found that Mn$_{0.5}$Fe$_{0.5}$PS$_3$ is an near-ideal example of a two-dimensional magnetic material through observing rod-like structures in single crystal data and determining that interactions between the planes were very weak in our analysis of the spin correlations.

\section{Data Availability}
\noindent
Raw data sets from ILL experiments can be accessed via links provided in Refs. 49 and 53. Magnetisation data presented in this paper resulting from the UK effort will be made available at https://wrap.warwick.ac.uk/132173.

\section{Acknowledgements}
\noindent
J. N. G. and L.C. acknowledge the ILL Graduate School and the EPSRC (DTP) for providing a studentship to J. N. G. This project has received funding from the European Research Council (ERC) under the European Union’s Horizon 2020 research and innovation programme (Grant Agreement No. 681260). This work was supported by the Institute for Basic Science (IBS) in Korea (Grant No. IBS-R009-G1). Work at the CQM was supported by the National Research Foundation of Korea (Grant No. 2020R1A3B2079375). The authors would like to thank Joe Paddison, Paul Goddard, Si$\mathrm{\hat{a}}$n Dutton, Charles Haines, David Jarvis, Cheng Liu, Siddharth Saxena, Inho Hwang, JungHyun Kim and Nahyun Lee for their help and useful discussions.
\bibliography{References}{}

\begin{thebibliography}{65}%
\makeatletter
\providecommand \@ifxundefined [1]{%
 \@ifx{#1\undefined}
}%
\providecommand \@ifnum [1]{%
 \ifnum #1\expandafter \@firstoftwo
 \else \expandafter \@secondoftwo
 \fi
}%
\providecommand \@ifx [1]{%
 \ifx #1\expandafter \@firstoftwo
 \else \expandafter \@secondoftwo
 \fi
}%
\providecommand \natexlab [1]{#1}%
\providecommand \enquote  [1]{``#1''}%
\providecommand \bibnamefont  [1]{#1}%
\providecommand \bibfnamefont [1]{#1}%
\providecommand \citenamefont [1]{#1}%
\providecommand \href@noop [0]{\@secondoftwo}%
\providecommand \href [0]{\begingroup \@sanitize@url \@href}%
\providecommand \@href[1]{\@@startlink{#1}\@@href}%
\providecommand \@@href[1]{\endgroup#1\@@endlink}%
\providecommand \@sanitize@url [0]{\catcode `\\12\catcode `\$12\catcode
  `\&12\catcode `\#12\catcode `\^12\catcode `\_12\catcode `\%12\relax}%
\providecommand \@@startlink[1]{}%
\providecommand \@@endlink[0]{}%
\providecommand \url  [0]{\begingroup\@sanitize@url \@url }%
\providecommand \@url [1]{\endgroup\@href {#1}{\urlprefix }}%
\providecommand \urlprefix  [0]{URL }%
\providecommand \Eprint [0]{\href }%
\providecommand \doibase [0]{http://dx.doi.org/}%
\providecommand \selectlanguage [0]{\@gobble}%
\providecommand \bibinfo  [0]{\@secondoftwo}%
\providecommand \bibfield  [0]{\@secondoftwo}%
\providecommand \translation [1]{[#1]}%
\providecommand \BibitemOpen [0]{}%
\providecommand \bibitemStop [0]{}%
\providecommand \bibitemNoStop [0]{.\EOS\space}%
\providecommand \EOS [0]{\spacefactor3000\relax}%
\providecommand \BibitemShut  [1]{\csname bibitem#1\endcsname}%
\let\auto@bib@innerbib\@empty
%</preamble>
\bibitem [{\citenamefont {Tang}\ and\ \citenamefont {Zhou}(2013)}]{Tang2013}%
  \BibitemOpen
  \bibfield  {author} {\bibinfo {author} {\bibfnamefont {Q.}~\bibnamefont
  {Tang}}\ and\ \bibinfo {author} {\bibfnamefont {Z.}~\bibnamefont {Zhou}},\
  }\href {\doibase 10.1016/j.pmatsci.2013.04.003} {\bibfield  {journal}
  {\bibinfo  {journal} {Prog. Mater. Sci.}\ }\textbf {\bibinfo {volume} {58}},\
  \bibinfo {pages} {1244} (\bibinfo {year} {2013})}\BibitemShut {NoStop}%
\bibitem [{\citenamefont {Wang}\ \emph {et~al.}(2018)\citenamefont {Wang},
  \citenamefont {Shifa}, \citenamefont {Yu}, \citenamefont {He}, \citenamefont
  {Liu}, \citenamefont {Wang}, \citenamefont {Wang}, \citenamefont {Zhan},
  \citenamefont {Lou}, \citenamefont {Xia},\ and\ \citenamefont
  {He}}]{Wang2018}%
  \BibitemOpen
  \bibfield  {author} {\bibinfo {author} {\bibfnamefont {F.}~\bibnamefont
  {Wang}}, \bibinfo {author} {\bibfnamefont {T.~A.}\ \bibnamefont {Shifa}},
  \bibinfo {author} {\bibfnamefont {P.}~\bibnamefont {Yu}}, \bibinfo {author}
  {\bibfnamefont {P.}~\bibnamefont {He}}, \bibinfo {author} {\bibfnamefont
  {Y.}~\bibnamefont {Liu}}, \bibinfo {author} {\bibfnamefont {F.}~\bibnamefont
  {Wang}}, \bibinfo {author} {\bibfnamefont {Z.}~\bibnamefont {Wang}}, \bibinfo
  {author} {\bibfnamefont {X.}~\bibnamefont {Zhan}}, \bibinfo {author}
  {\bibfnamefont {X.}~\bibnamefont {Lou}}, \bibinfo {author} {\bibfnamefont
  {F.}~\bibnamefont {Xia}}, \ and\ \bibinfo {author} {\bibfnamefont
  {J.}~\bibnamefont {He}},\ }\href {\doibase 10.1002/adfm.201802151} {\bibfield
   {journal} {\bibinfo  {journal} {Adv. Funct. Mater.}\ }\textbf {\bibinfo
  {volume} {28}},\ \bibinfo {pages} {1802151} (\bibinfo {year}
  {2018})}\BibitemShut {NoStop}%
\bibitem [{\citenamefont {Ponraj}\ \emph {et~al.}(2016)\citenamefont {Ponraj},
  \citenamefont {Xu}, \citenamefont {Dhanabalan}, \citenamefont {Mu},
  \citenamefont {Wang}, \citenamefont {Yuan}, \citenamefont {Li}, \citenamefont
  {Thakur}, \citenamefont {Ashrafi}, \citenamefont {Mccoubrey}, \citenamefont
  {Zhang}, \citenamefont {Li}, \citenamefont {Zhang},\ and\ \citenamefont
  {Bao}}]{Ponraj_2016}%
  \BibitemOpen
  \bibfield  {author} {\bibinfo {author} {\bibfnamefont {J.~S.}\ \bibnamefont
  {Ponraj}}, \bibinfo {author} {\bibfnamefont {Z.-Q.}\ \bibnamefont {Xu}},
  \bibinfo {author} {\bibfnamefont {S.~C.}\ \bibnamefont {Dhanabalan}},
  \bibinfo {author} {\bibfnamefont {H.}~\bibnamefont {Mu}}, \bibinfo {author}
  {\bibfnamefont {Y.}~\bibnamefont {Wang}}, \bibinfo {author} {\bibfnamefont
  {J.}~\bibnamefont {Yuan}}, \bibinfo {author} {\bibfnamefont {P.}~\bibnamefont
  {Li}}, \bibinfo {author} {\bibfnamefont {S.}~\bibnamefont {Thakur}}, \bibinfo
  {author} {\bibfnamefont {M.}~\bibnamefont {Ashrafi}}, \bibinfo {author}
  {\bibfnamefont {K.}~\bibnamefont {Mccoubrey}}, \bibinfo {author}
  {\bibfnamefont {Y.}~\bibnamefont {Zhang}}, \bibinfo {author} {\bibfnamefont
  {S.}~\bibnamefont {Li}}, \bibinfo {author} {\bibfnamefont {H.}~\bibnamefont
  {Zhang}}, \ and\ \bibinfo {author} {\bibfnamefont {Q.}~\bibnamefont {Bao}},\
  }\href {\doibase 10.1088/0957-4484/27/46/462001} {\bibfield  {journal}
  {\bibinfo  {journal} {Nanotechnology}\ }\textbf {\bibinfo {volume} {27}},\
  \bibinfo {pages} {462001} (\bibinfo {year} {2016})}\BibitemShut {NoStop}%
\bibitem [{\citenamefont {Mak}\ and\ \citenamefont {Shan}(2016)}]{Mak2016}%
  \BibitemOpen
  \bibfield  {author} {\bibinfo {author} {\bibfnamefont {K.~F.}\ \bibnamefont
  {Mak}}\ and\ \bibinfo {author} {\bibfnamefont {J.}~\bibnamefont {Shan}},\
  }\href {\doibase 10.1038/nphoton.2015.282} {\bibfield  {journal} {\bibinfo
  {journal} {Nat. Photonics}\ }\textbf {\bibinfo {volume} {10}},\ \bibinfo
  {pages} {216} (\bibinfo {year} {2016})}\BibitemShut {NoStop}%
\bibitem [{\citenamefont {Kumar}\ and\ \citenamefont {Xu}(2018)}]{Kumar2018}%
  \BibitemOpen
  \bibfield  {author} {\bibinfo {author} {\bibfnamefont {A.}~\bibnamefont
  {Kumar}}\ and\ \bibinfo {author} {\bibfnamefont {Q.}~\bibnamefont {Xu}},\
  }\href {\doibase 10.1002/cnma.201700139} {\bibfield  {journal} {\bibinfo
  {journal} {ChemNanoMat.}\ }\textbf {\bibinfo {volume} {4}},\ \bibinfo {pages}
  {28} (\bibinfo {year} {2018})}\BibitemShut {NoStop}%
\bibitem [{\citenamefont {Burch}\ \emph {et~al.}(2018)\citenamefont {Burch},
  \citenamefont {Mandrus},\ and\ \citenamefont {Park}}]{Burch2018}%
  \BibitemOpen
  \bibfield  {author} {\bibinfo {author} {\bibfnamefont {K.~S.}\ \bibnamefont
  {Burch}}, \bibinfo {author} {\bibfnamefont {D.}~\bibnamefont {Mandrus}}, \
  and\ \bibinfo {author} {\bibfnamefont {J.-G.}\ \bibnamefont {Park}},\ }\href
  {\doibase 10.1038/s41586-018-0631-z} {\bibfield  {journal} {\bibinfo
  {journal} {Nature}\ }\textbf {\bibinfo {volume} {563}},\ \bibinfo {pages}
  {47} (\bibinfo {year} {2018})}\BibitemShut {NoStop}%
\bibitem [{\citenamefont {Park}(2016)}]{Park2016}%
  \BibitemOpen
  \bibfield  {author} {\bibinfo {author} {\bibfnamefont {J.-G.}\ \bibnamefont
  {Park}},\ }\href@noop {} {\bibfield  {journal} {\bibinfo  {journal} {J. Phys.
  Condens. Matter}\ }\textbf {\bibinfo {volume} {28}},\ \bibinfo {pages} {30}
  (\bibinfo {year} {2016})}\BibitemShut {NoStop}%
\bibitem [{\citenamefont {Dikin}\ \emph {et~al.}(2011)\citenamefont {Dikin},
  \citenamefont {Mehta}, \citenamefont {Bark}, \citenamefont {Folkman},
  \citenamefont {Eom},\ and\ \citenamefont
  {Chandrasekhar}}]{PhysRevLett.107.056802}%
  \BibitemOpen
  \bibfield  {author} {\bibinfo {author} {\bibfnamefont {D.~A.}\ \bibnamefont
  {Dikin}}, \bibinfo {author} {\bibfnamefont {M.}~\bibnamefont {Mehta}},
  \bibinfo {author} {\bibfnamefont {C.~W.}\ \bibnamefont {Bark}}, \bibinfo
  {author} {\bibfnamefont {C.~M.}\ \bibnamefont {Folkman}}, \bibinfo {author}
  {\bibfnamefont {C.~B.}\ \bibnamefont {Eom}}, \ and\ \bibinfo {author}
  {\bibfnamefont {V.}~\bibnamefont {Chandrasekhar}},\ }\href {\doibase
  10.1103/PhysRevLett.107.056802} {\bibfield  {journal} {\bibinfo  {journal}
  {Phys. Rev. Lett.}\ }\textbf {\bibinfo {volume} {107}},\ \bibinfo {pages}
  {056802} (\bibinfo {year} {2011})}\BibitemShut {NoStop}%
\bibitem [{\citenamefont {Gong}\ \emph {et~al.}(2017)\citenamefont {Gong},
  \citenamefont {Li}, \citenamefont {Li}, \citenamefont {Ji}, \citenamefont
  {Stern}, \citenamefont {Xia}, \citenamefont {Cao}, \citenamefont {Bao},
  \citenamefont {Wang}, \citenamefont {Wang}, \citenamefont {Qiu},
  \citenamefont {Cava}, \citenamefont {Louie}, \citenamefont {Xia},\ and\
  \citenamefont {Zhang}}]{Gong2017}%
  \BibitemOpen
  \bibfield  {author} {\bibinfo {author} {\bibfnamefont {C.}~\bibnamefont
  {Gong}}, \bibinfo {author} {\bibfnamefont {L.}~\bibnamefont {Li}}, \bibinfo
  {author} {\bibfnamefont {Z.}~\bibnamefont {Li}}, \bibinfo {author}
  {\bibfnamefont {H.}~\bibnamefont {Ji}}, \bibinfo {author} {\bibfnamefont
  {A.}~\bibnamefont {Stern}}, \bibinfo {author} {\bibfnamefont
  {Y.}~\bibnamefont {Xia}}, \bibinfo {author} {\bibfnamefont {T.}~\bibnamefont
  {Cao}}, \bibinfo {author} {\bibfnamefont {W.}~\bibnamefont {Bao}}, \bibinfo
  {author} {\bibfnamefont {C.}~\bibnamefont {Wang}}, \bibinfo {author}
  {\bibfnamefont {Y.}~\bibnamefont {Wang}}, \bibinfo {author} {\bibfnamefont
  {Z.~Q.}\ \bibnamefont {Qiu}}, \bibinfo {author} {\bibfnamefont {R.~J.}\
  \bibnamefont {Cava}}, \bibinfo {author} {\bibfnamefont {S.~G.}\ \bibnamefont
  {Louie}}, \bibinfo {author} {\bibfnamefont {J.}~\bibnamefont {Xia}}, \ and\
  \bibinfo {author} {\bibfnamefont {X.}~\bibnamefont {Zhang}},\ }\href
  {\doibase 10.1038/nature22060} {\bibfield  {journal} {\bibinfo  {journal}
  {Nature}\ }\textbf {\bibinfo {volume} {546}},\ \bibinfo {pages} {265}
  (\bibinfo {year} {2017})}\BibitemShut {NoStop}%
\bibitem [{\citenamefont {Wang}\ \emph {et~al.}(2016)\citenamefont {Wang},
  \citenamefont {Fan}, \citenamefont {Zhu},\ and\ \citenamefont
  {Wu}}]{Wang_2016}%
  \BibitemOpen
  \bibfield  {author} {\bibinfo {author} {\bibfnamefont {H.}~\bibnamefont
  {Wang}}, \bibinfo {author} {\bibfnamefont {F.}~\bibnamefont {Fan}}, \bibinfo
  {author} {\bibfnamefont {S.}~\bibnamefont {Zhu}}, \ and\ \bibinfo {author}
  {\bibfnamefont {H.}~\bibnamefont {Wu}},\ }\href {\doibase
  10.1209/0295-5075/114/47001} {\bibfield  {journal} {\bibinfo  {journal} {EPL-
  EUROPHYS LETT}\ }\textbf {\bibinfo {volume} {114}},\ \bibinfo {pages} {47001}
  (\bibinfo {year} {2016})}\BibitemShut {NoStop}%
\bibitem [{\citenamefont {Huang}\ \emph {et~al.}(2018)\citenamefont {Huang},
  \citenamefont {Clark}, \citenamefont {Klein}, \citenamefont {MacNeill},
  \citenamefont {Navarro-Moratalla}, \citenamefont {Seyler}, \citenamefont
  {Wilson}, \citenamefont {McGuire}, \citenamefont {Cobden}, \citenamefont
  {Xiao}, \citenamefont {Yao}, \citenamefont {Jarillo-Herrero},\ and\
  \citenamefont {Xu}}]{Huang2018}%
  \BibitemOpen
  \bibfield  {author} {\bibinfo {author} {\bibfnamefont {B.}~\bibnamefont
  {Huang}}, \bibinfo {author} {\bibfnamefont {G.}~\bibnamefont {Clark}},
  \bibinfo {author} {\bibfnamefont {D.~R.}\ \bibnamefont {Klein}}, \bibinfo
  {author} {\bibfnamefont {D.}~\bibnamefont {MacNeill}}, \bibinfo {author}
  {\bibfnamefont {E.}~\bibnamefont {Navarro-Moratalla}}, \bibinfo {author}
  {\bibfnamefont {K.~L.}\ \bibnamefont {Seyler}}, \bibinfo {author}
  {\bibfnamefont {N.}~\bibnamefont {Wilson}}, \bibinfo {author} {\bibfnamefont
  {M.~A.}\ \bibnamefont {McGuire}}, \bibinfo {author} {\bibfnamefont {D.~H.}\
  \bibnamefont {Cobden}}, \bibinfo {author} {\bibfnamefont {D.}~\bibnamefont
  {Xiao}}, \bibinfo {author} {\bibfnamefont {W.}~\bibnamefont {Yao}}, \bibinfo
  {author} {\bibfnamefont {P.}~\bibnamefont {Jarillo-Herrero}}, \ and\ \bibinfo
  {author} {\bibfnamefont {X.}~\bibnamefont {Xu}},\ }\href {\doibase
  10.1038/s41565-018-0121-3} {\bibfield  {journal} {\bibinfo  {journal} {Nat.
  Nanotechnol.}\ }\textbf {\bibinfo {volume} {13}},\ \bibinfo {pages} {544}
  (\bibinfo {year} {2018})},\ \Eprint {http://arxiv.org/abs/1802.06979}
  {1802.06979} \BibitemShut {NoStop}%
\bibitem [{\citenamefont {Coak}\ \emph {et~al.}(2020)\citenamefont {Coak},
  \citenamefont {Jarvis}, \citenamefont {Hamidov}, \citenamefont {Haines},
  \citenamefont {Alireza}, \citenamefont {Liu}, \citenamefont {Son},
  \citenamefont {Hwang}, \citenamefont {Lampronti}, \citenamefont
  {Daisenberger}, \citenamefont {Nahai-Williamson}, \citenamefont {Wildes},
  \citenamefont {Saxena},\ and\ \citenamefont {Park}}]{Coak2020}%
  \BibitemOpen
  \bibfield  {author} {\bibinfo {author} {\bibfnamefont {M.~J.}\ \bibnamefont
  {Coak}}, \bibinfo {author} {\bibfnamefont {D.~M.}\ \bibnamefont {Jarvis}},
  \bibinfo {author} {\bibfnamefont {H.}~\bibnamefont {Hamidov}}, \bibinfo
  {author} {\bibfnamefont {C.~R.~S.}\ \bibnamefont {Haines}}, \bibinfo {author}
  {\bibfnamefont {P.~L.}\ \bibnamefont {Alireza}}, \bibinfo {author}
  {\bibfnamefont {C.}~\bibnamefont {Liu}}, \bibinfo {author} {\bibfnamefont
  {S.}~\bibnamefont {Son}}, \bibinfo {author} {\bibfnamefont {I.}~\bibnamefont
  {Hwang}}, \bibinfo {author} {\bibfnamefont {G.~I.}\ \bibnamefont
  {Lampronti}}, \bibinfo {author} {\bibfnamefont {D.}~\bibnamefont
  {Daisenberger}}, \bibinfo {author} {\bibfnamefont {P.}~\bibnamefont
  {Nahai-Williamson}}, \bibinfo {author} {\bibfnamefont {A.~R.}\ \bibnamefont
  {Wildes}}, \bibinfo {author} {\bibfnamefont {S.~S.}\ \bibnamefont {Saxena}},
  \ and\ \bibinfo {author} {\bibfnamefont {J.-G.}\ \bibnamefont {Park}},\
  }\href@noop {} {\bibfield  {journal} {\bibinfo  {journal} {J. Phys. Condens.
  Matter}\ }\textbf {\bibinfo {volume} {32}},\ \bibinfo {pages} {124003}
  (\bibinfo {year} {2020})}\BibitemShut {NoStop}%
\bibitem [{\citenamefont {Haines}\ \emph {et~al.}(2018)\citenamefont {Haines},
  \citenamefont {Coak}, \citenamefont {Wildes}, \citenamefont {Lampronti},
  \citenamefont {Liu}, \citenamefont {Nahai-Williamson}, \citenamefont
  {Hamidov}, \citenamefont {Daisenberger},\ and\ \citenamefont
  {Saxena}}]{Haines2018}%
  \BibitemOpen
  \bibfield  {author} {\bibinfo {author} {\bibfnamefont {C.~R.~S.}\
  \bibnamefont {Haines}}, \bibinfo {author} {\bibfnamefont {M.~J.}\
  \bibnamefont {Coak}}, \bibinfo {author} {\bibfnamefont {A.~R.}\ \bibnamefont
  {Wildes}}, \bibinfo {author} {\bibfnamefont {G.~I.}\ \bibnamefont
  {Lampronti}}, \bibinfo {author} {\bibfnamefont {C.}~\bibnamefont {Liu}},
  \bibinfo {author} {\bibfnamefont {P.}~\bibnamefont {Nahai-Williamson}},
  \bibinfo {author} {\bibfnamefont {H.}~\bibnamefont {Hamidov}}, \bibinfo
  {author} {\bibfnamefont {D.}~\bibnamefont {Daisenberger}}, \ and\ \bibinfo
  {author} {\bibfnamefont {S.~S.}\ \bibnamefont {Saxena}},\ }\href {\doibase
  10.1103/PhysRevLett.121.266801} {\bibfield  {journal} {\bibinfo  {journal}
  {Phys. Rev. Lett.}\ }\textbf {\bibinfo {volume} {121}},\ \bibinfo {pages}
  {266801} (\bibinfo {year} {2018})}\BibitemShut {NoStop}%
\bibitem [{\citenamefont {Balents}(2010)}]{Balents2010}%
  \BibitemOpen
  \bibfield  {author} {\bibinfo {author} {\bibfnamefont {L.}~\bibnamefont
  {Balents}},\ }\href@noop {} {\bibfield  {journal} {\bibinfo  {journal}
  {Nature}\ }\textbf {\bibinfo {volume} {464}},\ \bibinfo {pages} {199}
  (\bibinfo {year} {2010})}\BibitemShut {NoStop}%
\bibitem [{\citenamefont {Savary}\ and\ \citenamefont
  {Balents}(2017)}]{Savary}%
  \BibitemOpen
  \bibfield  {author} {\bibinfo {author} {\bibfnamefont {L.}~\bibnamefont
  {Savary}}\ and\ \bibinfo {author} {\bibfnamefont {L.}~\bibnamefont
  {Balents}},\ }\href {\doibase 10.1088/0034-4885/80/1/016502} {\bibfield
  {journal} {\bibinfo  {journal} {Rep. Prog. Phys.}\ }\textbf {\bibinfo
  {volume} {80}},\ \bibinfo {pages} {016502} (\bibinfo {year}
  {2017})}\BibitemShut {NoStop}%
\bibitem [{\citenamefont {Kitaev}(2006)}]{Kitaev}%
  \BibitemOpen
  \bibfield  {author} {\bibinfo {author} {\bibfnamefont {A.}~\bibnamefont
  {Kitaev}},\ }\href {\doibase 10.1016/j.aop.2005.10.005} {\bibfield  {journal}
  {\bibinfo  {journal} {Ann. Phys.}\ }\textbf {\bibinfo {volume} {321}},\
  \bibinfo {pages} {2} (\bibinfo {year} {2006})}\BibitemShut {NoStop}%
\bibitem [{\citenamefont {Hermanns}\ \emph {et~al.}(2018)\citenamefont
  {Hermanns}, \citenamefont {Kimchi},\ and\ \citenamefont {Knolle}}]{kitaev2}%
  \BibitemOpen
  \bibfield  {author} {\bibinfo {author} {\bibfnamefont {M.}~\bibnamefont
  {Hermanns}}, \bibinfo {author} {\bibfnamefont {I.}~\bibnamefont {Kimchi}}, \
  and\ \bibinfo {author} {\bibfnamefont {J.}~\bibnamefont {Knolle}},\ }\href
  {\doibase 10.1146/annurev-conmatphys-033117-053934} {\bibfield  {journal}
  {\bibinfo  {journal} {Annu. Rev. Condens. Matter Phys.}\ }\textbf {\bibinfo
  {volume} {9}},\ \bibinfo {pages} {1} (\bibinfo {year} {2018})}\BibitemShut
  {NoStop}%
\bibitem [{\citenamefont {Banerjee}\ \emph {et~al.}(2016)\citenamefont
  {Banerjee}, \citenamefont {Bridges}, \citenamefont {Yan}, \citenamefont
  {Aczel}, \citenamefont {Li}, \citenamefont {Stone}, \citenamefont {Granroth},
  \citenamefont {Lumsden}, \citenamefont {Yiu}, \citenamefont {Knolle},
  \citenamefont {Bhattacharjee}, \citenamefont {Kovrizhin}, \citenamefont
  {Moessner}, \citenamefont {Tennant}, \citenamefont {Mandrus},\ and\
  \citenamefont {Nagler}}]{Banerjee2016}%
  \BibitemOpen
  \bibfield  {author} {\bibinfo {author} {\bibfnamefont {A.}~\bibnamefont
  {Banerjee}}, \bibinfo {author} {\bibfnamefont {C.~A.}\ \bibnamefont
  {Bridges}}, \bibinfo {author} {\bibfnamefont {J.-Q.}\ \bibnamefont {Yan}},
  \bibinfo {author} {\bibfnamefont {A.~A.}\ \bibnamefont {Aczel}}, \bibinfo
  {author} {\bibfnamefont {L.}~\bibnamefont {Li}}, \bibinfo {author}
  {\bibfnamefont {M.~B.}\ \bibnamefont {Stone}}, \bibinfo {author}
  {\bibfnamefont {G.~E.}\ \bibnamefont {Granroth}}, \bibinfo {author}
  {\bibfnamefont {M.~D.}\ \bibnamefont {Lumsden}}, \bibinfo {author}
  {\bibfnamefont {Y.}~\bibnamefont {Yiu}}, \bibinfo {author} {\bibfnamefont
  {J.}~\bibnamefont {Knolle}}, \bibinfo {author} {\bibfnamefont
  {S.}~\bibnamefont {Bhattacharjee}}, \bibinfo {author} {\bibfnamefont {D.~L.}\
  \bibnamefont {Kovrizhin}}, \bibinfo {author} {\bibfnamefont {R.}~\bibnamefont
  {Moessner}}, \bibinfo {author} {\bibfnamefont {D.~A.}\ \bibnamefont
  {Tennant}}, \bibinfo {author} {\bibfnamefont {D.~G.}\ \bibnamefont
  {Mandrus}}, \ and\ \bibinfo {author} {\bibfnamefont {S.~E.}\ \bibnamefont
  {Nagler}},\ }\href {\doibase 10.1038/nmat4604} {\bibfield  {journal}
  {\bibinfo  {journal} {Nat. Mater.}\ }\textbf {\bibinfo {volume} {15}},\
  \bibinfo {pages} {733} (\bibinfo {year} {2016})}\BibitemShut {NoStop}%
\bibitem [{\citenamefont {Kim}\ \emph {et~al.}(2015)\citenamefont {Kim},
  \citenamefont {Vijay~Shankar}, \citenamefont {Catuneanu},\ and\ \citenamefont
  {Kee}}]{PhysRevB.91.241110}%
  \BibitemOpen
  \bibfield  {author} {\bibinfo {author} {\bibfnamefont {H.-S.}\ \bibnamefont
  {Kim}}, \bibinfo {author} {\bibfnamefont {V.}~\bibnamefont {Vijay~Shankar}},
  \bibinfo {author} {\bibfnamefont {A.}~\bibnamefont {Catuneanu}}, \ and\
  \bibinfo {author} {\bibfnamefont {H.-Y.}\ \bibnamefont {Kee}},\ }\href
  {\doibase 10.1103/PhysRevB.91.241110} {\bibfield  {journal} {\bibinfo
  {journal} {Phys. Rev. B}\ }\textbf {\bibinfo {volume} {91}},\ \bibinfo
  {pages} {241110(R)} (\bibinfo {year} {2015})}\BibitemShut {NoStop}%
\bibitem [{\citenamefont {Sears}\ \emph {et~al.}(2015)\citenamefont {Sears},
  \citenamefont {Songvilay}, \citenamefont {Plumb}, \citenamefont {Clancy},
  \citenamefont {Qiu}, \citenamefont {Zhao}, \citenamefont {Parshall},\ and\
  \citenamefont {Kim}}]{PhysRevB.91.144420}%
  \BibitemOpen
  \bibfield  {author} {\bibinfo {author} {\bibfnamefont {J.~A.}\ \bibnamefont
  {Sears}}, \bibinfo {author} {\bibfnamefont {M.}~\bibnamefont {Songvilay}},
  \bibinfo {author} {\bibfnamefont {K.~W.}\ \bibnamefont {Plumb}}, \bibinfo
  {author} {\bibfnamefont {J.~P.}\ \bibnamefont {Clancy}}, \bibinfo {author}
  {\bibfnamefont {Y.}~\bibnamefont {Qiu}}, \bibinfo {author} {\bibfnamefont
  {Y.}~\bibnamefont {Zhao}}, \bibinfo {author} {\bibfnamefont {D.}~\bibnamefont
  {Parshall}}, \ and\ \bibinfo {author} {\bibfnamefont {Y.-J.}\ \bibnamefont
  {Kim}},\ }\href {\doibase 10.1103/PhysRevB.91.144420} {\bibfield  {journal}
  {\bibinfo  {journal} {Phys. Rev. B}\ }\textbf {\bibinfo {volume} {91}},\
  \bibinfo {pages} {144420} (\bibinfo {year} {2015})}\BibitemShut {NoStop}%
\bibitem [{\citenamefont {Sandilands}\ \emph {et~al.}(2016)\citenamefont
  {Sandilands}, \citenamefont {Tian}, \citenamefont {Reijnders}, \citenamefont
  {Kim}, \citenamefont {Plumb}, \citenamefont {Kim}, \citenamefont {Kee},\ and\
  \citenamefont {Burch}}]{PhysRevB.93.075144}%
  \BibitemOpen
  \bibfield  {author} {\bibinfo {author} {\bibfnamefont {L.~J.}\ \bibnamefont
  {Sandilands}}, \bibinfo {author} {\bibfnamefont {Y.}~\bibnamefont {Tian}},
  \bibinfo {author} {\bibfnamefont {A.~A.}\ \bibnamefont {Reijnders}}, \bibinfo
  {author} {\bibfnamefont {H.-S.}\ \bibnamefont {Kim}}, \bibinfo {author}
  {\bibfnamefont {K.~W.}\ \bibnamefont {Plumb}}, \bibinfo {author}
  {\bibfnamefont {Y.-J.}\ \bibnamefont {Kim}}, \bibinfo {author} {\bibfnamefont
  {H.-Y.}\ \bibnamefont {Kee}}, \ and\ \bibinfo {author} {\bibfnamefont
  {K.~S.}\ \bibnamefont {Burch}},\ }\href {\doibase 10.1103/PhysRevB.93.075144}
  {\bibfield  {journal} {\bibinfo  {journal} {Phys. Rev. B}\ }\textbf {\bibinfo
  {volume} {93}},\ \bibinfo {pages} {075144} (\bibinfo {year}
  {2016})}\BibitemShut {NoStop}%
\bibitem [{\citenamefont {Williams}\ \emph {et~al.}(2016)\citenamefont
  {Williams}, \citenamefont {Johnson}, \citenamefont {Freund}, \citenamefont
  {Choi}, \citenamefont {Jesche}, \citenamefont {Kimchi}, \citenamefont
  {Manni}, \citenamefont {Bombardi}, \citenamefont {Manuel}, \citenamefont
  {Gegenwart},\ and\ \citenamefont {Coldea}}]{PhysRevB.93.195158}%
  \BibitemOpen
  \bibfield  {author} {\bibinfo {author} {\bibfnamefont {S.~C.}\ \bibnamefont
  {Williams}}, \bibinfo {author} {\bibfnamefont {R.~D.}\ \bibnamefont
  {Johnson}}, \bibinfo {author} {\bibfnamefont {F.}~\bibnamefont {Freund}},
  \bibinfo {author} {\bibfnamefont {S.}~\bibnamefont {Choi}}, \bibinfo {author}
  {\bibfnamefont {A.}~\bibnamefont {Jesche}}, \bibinfo {author} {\bibfnamefont
  {I.}~\bibnamefont {Kimchi}}, \bibinfo {author} {\bibfnamefont
  {S.}~\bibnamefont {Manni}}, \bibinfo {author} {\bibfnamefont
  {A.}~\bibnamefont {Bombardi}}, \bibinfo {author} {\bibfnamefont
  {P.}~\bibnamefont {Manuel}}, \bibinfo {author} {\bibfnamefont
  {P.}~\bibnamefont {Gegenwart}}, \ and\ \bibinfo {author} {\bibfnamefont
  {R.}~\bibnamefont {Coldea}},\ }\href {\doibase 10.1103/PhysRevB.93.195158}
  {\bibfield  {journal} {\bibinfo  {journal} {Phys. Rev. B}\ }\textbf {\bibinfo
  {volume} {93}},\ \bibinfo {pages} {195158} (\bibinfo {year}
  {2016})}\BibitemShut {NoStop}%
\bibitem [{\citenamefont {Ye}\ \emph {et~al.}(2012)\citenamefont {Ye},
  \citenamefont {Chi}, \citenamefont {Cao}, \citenamefont {Chakoumakos},
  \citenamefont {Fernandez-Baca}, \citenamefont {Custelcean}, \citenamefont
  {Qi}, \citenamefont {Korneta},\ and\ \citenamefont
  {Cao}}]{PhysRevB.85.180403}%
  \BibitemOpen
  \bibfield  {author} {\bibinfo {author} {\bibfnamefont {F.}~\bibnamefont
  {Ye}}, \bibinfo {author} {\bibfnamefont {S.}~\bibnamefont {Chi}}, \bibinfo
  {author} {\bibfnamefont {H.}~\bibnamefont {Cao}}, \bibinfo {author}
  {\bibfnamefont {B.~C.}\ \bibnamefont {Chakoumakos}}, \bibinfo {author}
  {\bibfnamefont {J.~A.}\ \bibnamefont {Fernandez-Baca}}, \bibinfo {author}
  {\bibfnamefont {R.}~\bibnamefont {Custelcean}}, \bibinfo {author}
  {\bibfnamefont {T.~F.}\ \bibnamefont {Qi}}, \bibinfo {author} {\bibfnamefont
  {O.~B.}\ \bibnamefont {Korneta}}, \ and\ \bibinfo {author} {\bibfnamefont
  {G.}~\bibnamefont {Cao}},\ }\href {\doibase 10.1103/PhysRevB.85.180403}
  {\bibfield  {journal} {\bibinfo  {journal} {Phys. Rev. B}\ }\textbf {\bibinfo
  {volume} {85}},\ \bibinfo {pages} {180403(R)} (\bibinfo {year}
  {2012})}\BibitemShut {NoStop}%
\bibitem [{\citenamefont {Liu}\ \emph {et~al.}(2011)\citenamefont {Liu},
  \citenamefont {Berlijn}, \citenamefont {Yin}, \citenamefont {Ku},
  \citenamefont {Tsvelik}, \citenamefont {Kim}, \citenamefont {Gretarsson},
  \citenamefont {Singh}, \citenamefont {Gegenwart},\ and\ \citenamefont
  {Hill}}]{PhysRevB.83.220403}%
  \BibitemOpen
  \bibfield  {author} {\bibinfo {author} {\bibfnamefont {X.}~\bibnamefont
  {Liu}}, \bibinfo {author} {\bibfnamefont {T.}~\bibnamefont {Berlijn}},
  \bibinfo {author} {\bibfnamefont {W.-G.}\ \bibnamefont {Yin}}, \bibinfo
  {author} {\bibfnamefont {W.}~\bibnamefont {Ku}}, \bibinfo {author}
  {\bibfnamefont {A.}~\bibnamefont {Tsvelik}}, \bibinfo {author} {\bibfnamefont
  {Y.-J.}\ \bibnamefont {Kim}}, \bibinfo {author} {\bibfnamefont
  {H.}~\bibnamefont {Gretarsson}}, \bibinfo {author} {\bibfnamefont
  {Y.}~\bibnamefont {Singh}}, \bibinfo {author} {\bibfnamefont
  {P.}~\bibnamefont {Gegenwart}}, \ and\ \bibinfo {author} {\bibfnamefont
  {J.~P.}\ \bibnamefont {Hill}},\ }\href {\doibase 10.1103/PhysRevB.83.220403}
  {\bibfield  {journal} {\bibinfo  {journal} {Phys. Rev. B}\ }\textbf {\bibinfo
  {volume} {83}},\ \bibinfo {pages} {220403(R)} (\bibinfo {year}
  {2011})}\BibitemShut {NoStop}%
\bibitem [{\citenamefont {Alaei}\ \emph {et~al.}(2017)\citenamefont {Alaei},
  \citenamefont {Mosadeq}, \citenamefont {Sarsari},\ and\ \citenamefont
  {Shahbazi}}]{PhysRevB.96.140404}%
  \BibitemOpen
  \bibfield  {author} {\bibinfo {author} {\bibfnamefont {M.}~\bibnamefont
  {Alaei}}, \bibinfo {author} {\bibfnamefont {H.}~\bibnamefont {Mosadeq}},
  \bibinfo {author} {\bibfnamefont {I.~A.}\ \bibnamefont {Sarsari}}, \ and\
  \bibinfo {author} {\bibfnamefont {F.}~\bibnamefont {Shahbazi}},\ }\href
  {\doibase 10.1103/PhysRevB.96.140404} {\bibfield  {journal} {\bibinfo
  {journal} {Phys. Rev. B}\ }\textbf {\bibinfo {volume} {96}},\ \bibinfo
  {pages} {140404(R)} (\bibinfo {year} {2017})}\BibitemShut {NoStop}%
\bibitem [{\citenamefont {Onishi}\ \emph {et~al.}(2012)\citenamefont {Onishi},
  \citenamefont {Oka}, \citenamefont {Azuma}, \citenamefont {Shimakawa},
  \citenamefont {Motome}, \citenamefont {Taniguchi}, \citenamefont {Hiraishi},
  \citenamefont {Miyazaki}, \citenamefont {Masuda}, \citenamefont {Koda},
  \citenamefont {Kojima},\ and\ \citenamefont {Kadono}}]{PhysRevB.85.184412}%
  \BibitemOpen
  \bibfield  {author} {\bibinfo {author} {\bibfnamefont {N.}~\bibnamefont
  {Onishi}}, \bibinfo {author} {\bibfnamefont {K.}~\bibnamefont {Oka}},
  \bibinfo {author} {\bibfnamefont {M.}~\bibnamefont {Azuma}}, \bibinfo
  {author} {\bibfnamefont {Y.}~\bibnamefont {Shimakawa}}, \bibinfo {author}
  {\bibfnamefont {Y.}~\bibnamefont {Motome}}, \bibinfo {author} {\bibfnamefont
  {T.}~\bibnamefont {Taniguchi}}, \bibinfo {author} {\bibfnamefont
  {M.}~\bibnamefont {Hiraishi}}, \bibinfo {author} {\bibfnamefont
  {M.}~\bibnamefont {Miyazaki}}, \bibinfo {author} {\bibfnamefont
  {T.}~\bibnamefont {Masuda}}, \bibinfo {author} {\bibfnamefont
  {A.}~\bibnamefont {Koda}}, \bibinfo {author} {\bibfnamefont {K.~M.}\
  \bibnamefont {Kojima}}, \ and\ \bibinfo {author} {\bibfnamefont
  {R.}~\bibnamefont {Kadono}},\ }\href {\doibase 10.1103/PhysRevB.85.184412}
  {\bibfield  {journal} {\bibinfo  {journal} {Phys. Rev. B}\ }\textbf {\bibinfo
  {volume} {85}},\ \bibinfo {pages} {184412} (\bibinfo {year}
  {2012})}\BibitemShut {NoStop}%
\bibitem [{\citenamefont {Matsuda}\ \emph {et~al.}(2010)\citenamefont
  {Matsuda}, \citenamefont {Azuma}, \citenamefont {Tokunaga}, \citenamefont
  {Shimakawa},\ and\ \citenamefont {Kumada}}]{PhysRevLett.105.187201}%
  \BibitemOpen
  \bibfield  {author} {\bibinfo {author} {\bibfnamefont {M.}~\bibnamefont
  {Matsuda}}, \bibinfo {author} {\bibfnamefont {M.}~\bibnamefont {Azuma}},
  \bibinfo {author} {\bibfnamefont {M.}~\bibnamefont {Tokunaga}}, \bibinfo
  {author} {\bibfnamefont {Y.}~\bibnamefont {Shimakawa}}, \ and\ \bibinfo
  {author} {\bibfnamefont {N.}~\bibnamefont {Kumada}},\ }\href {\doibase
  10.1103/PhysRevLett.105.187201} {\bibfield  {journal} {\bibinfo  {journal}
  {Phys. Rev. Lett.}\ }\textbf {\bibinfo {volume} {105}},\ \bibinfo {pages}
  {187201} (\bibinfo {year} {2010})}\BibitemShut {NoStop}%
\bibitem [{\citenamefont {Okumura}\ \emph {et~al.}(2010)\citenamefont
  {Okumura}, \citenamefont {Kawamura}, \citenamefont {Okubo},\ and\
  \citenamefont {Motome}}]{Okumura10}%
  \BibitemOpen
  \bibfield  {author} {\bibinfo {author} {\bibfnamefont {S.}~\bibnamefont
  {Okumura}}, \bibinfo {author} {\bibfnamefont {H.}~\bibnamefont {Kawamura}},
  \bibinfo {author} {\bibfnamefont {T.}~\bibnamefont {Okubo}}, \ and\ \bibinfo
  {author} {\bibfnamefont {Y.}~\bibnamefont {Motome}},\ }\href@noop {}
  {\bibfield  {journal} {\bibinfo  {journal} {J. Phys. Soc. Jpn.}\ }\textbf
  {\bibinfo {volume} {79}} (\bibinfo {year} {2010})}\BibitemShut {NoStop}%
\bibitem [{\citenamefont {Momma}\ and\ \citenamefont {Izumi}(2011)}]{Vesta}%
  \BibitemOpen
  \bibfield  {author} {\bibinfo {author} {\bibfnamefont {K.}~\bibnamefont
  {Momma}}\ and\ \bibinfo {author} {\bibfnamefont {F.}~\bibnamefont {Izumi}},\
  }\href@noop {} {\bibfield  {journal} {\bibinfo  {journal} {J. Appl. Cryst.}\
  }\textbf {\bibinfo {volume} {44}},\ \bibinfo {pages} {1272} (\bibinfo {year}
  {2011})}\BibitemShut {NoStop}%
\bibitem [{\citenamefont {Ouvrard}\ \emph {et~al.}(1985)\citenamefont
  {Ouvrard}, \citenamefont {Brec},\ and\ \citenamefont {Rouxel}}]{Ouvrard1985}%
  \BibitemOpen
  \bibfield  {author} {\bibinfo {author} {\bibfnamefont {G.}~\bibnamefont
  {Ouvrard}}, \bibinfo {author} {\bibfnamefont {R.}~\bibnamefont {Brec}}, \
  and\ \bibinfo {author} {\bibfnamefont {J.}~\bibnamefont {Rouxel}},\ }\href
  {\doibase 10.1016/0025-5408(85)90092-3} {\bibfield  {journal} {\bibinfo
  {journal} {Mater. Res. Bull.}\ }\textbf {\bibinfo {volume} {20}},\ \bibinfo
  {pages} {1181} (\bibinfo {year} {1985})}\BibitemShut {NoStop}%
\bibitem [{\citenamefont {Brec}\ \emph {et~al.}(1979)\citenamefont {Brec},
  \citenamefont {Schleich}, \citenamefont {Ouvrard}, \citenamefont {Louisy},\
  and\ \citenamefont {Rouxel}}]{Brec1979}%
  \BibitemOpen
  \bibfield  {author} {\bibinfo {author} {\bibfnamefont {R.}~\bibnamefont
  {Brec}}, \bibinfo {author} {\bibfnamefont {D.~M.}\ \bibnamefont {Schleich}},
  \bibinfo {author} {\bibfnamefont {G.}~\bibnamefont {Ouvrard}}, \bibinfo
  {author} {\bibfnamefont {A.}~\bibnamefont {Louisy}}, \ and\ \bibinfo {author}
  {\bibfnamefont {J.}~\bibnamefont {Rouxel}},\ }\href {\doibase
  10.1021/ic50197a018} {\bibfield  {journal} {\bibinfo  {journal} {Inorg.
  Chem.}\ }\textbf {\bibinfo {volume} {18}},\ \bibinfo {pages} {1814} (\bibinfo
  {year} {1979})}\BibitemShut {NoStop}%
\bibitem [{\citenamefont {Foot}\ \emph {et~al.}(1987)\citenamefont {Foot},
  \citenamefont {Katz}, \citenamefont {Patel}, \citenamefont {Nevett},
  \citenamefont {Pieecy},\ and\ \citenamefont {Balchin}}]{Foot1987}%
  \BibitemOpen
  \bibfield  {author} {\bibinfo {author} {\bibfnamefont {P.~J.~S.}\
  \bibnamefont {Foot}}, \bibinfo {author} {\bibfnamefont {T.}~\bibnamefont
  {Katz}}, \bibinfo {author} {\bibfnamefont {S.~N.}\ \bibnamefont {Patel}},
  \bibinfo {author} {\bibfnamefont {B.~A.}\ \bibnamefont {Nevett}}, \bibinfo
  {author} {\bibfnamefont {A.~R.}\ \bibnamefont {Pieecy}}, \ and\ \bibinfo
  {author} {\bibfnamefont {A.~A.}\ \bibnamefont {Balchin}},\ }\href {\doibase
  10.1002/pssa.2211000102} {\bibfield  {journal} {\bibinfo  {journal} {Phys.
  Status Solidi A}\ }\textbf {\bibinfo {volume} {100}},\ \bibinfo {pages} {11}
  (\bibinfo {year} {1987})}\BibitemShut {NoStop}%
\bibitem [{\citenamefont {Grasso}\ and\ \citenamefont
  {Silipigni}(2002)}]{Grasso}%
  \BibitemOpen
  \bibfield  {author} {\bibinfo {author} {\bibfnamefont {V.}~\bibnamefont
  {Grasso}}\ and\ \bibinfo {author} {\bibfnamefont {L.}~\bibnamefont
  {Silipigni}},\ }\href@noop {} {\bibfield  {journal} {\bibinfo  {journal}
  {Riv. Nuovo Cimento}\ }\textbf {\bibinfo {volume} {25}},\ \bibinfo {pages}
  {1} (\bibinfo {year} {2002})}\BibitemShut {NoStop}%
\bibitem [{\citenamefont {Kuo}\ \emph {et~al.}(2016)\citenamefont {Kuo},
  \citenamefont {Neumann}, \citenamefont {Balamurugan}, \citenamefont {Park},
  \citenamefont {Kang}, \citenamefont {Shiu}, \citenamefont {Kang},
  \citenamefont {Hong}, \citenamefont {Han}, \citenamefont {Noh},\ and\
  \citenamefont {Park}}]{Kuo2016}%
  \BibitemOpen
  \bibfield  {author} {\bibinfo {author} {\bibfnamefont {C.-T.}\ \bibnamefont
  {Kuo}}, \bibinfo {author} {\bibfnamefont {M.}~\bibnamefont {Neumann}},
  \bibinfo {author} {\bibfnamefont {K.}~\bibnamefont {Balamurugan}}, \bibinfo
  {author} {\bibfnamefont {H.~J.}\ \bibnamefont {Park}}, \bibinfo {author}
  {\bibfnamefont {S.}~\bibnamefont {Kang}}, \bibinfo {author} {\bibfnamefont
  {H.~W.}\ \bibnamefont {Shiu}}, \bibinfo {author} {\bibfnamefont {J.~H.}\
  \bibnamefont {Kang}}, \bibinfo {author} {\bibfnamefont {B.~H.}\ \bibnamefont
  {Hong}}, \bibinfo {author} {\bibfnamefont {M.}~\bibnamefont {Han}}, \bibinfo
  {author} {\bibfnamefont {T.~W.}\ \bibnamefont {Noh}}, \ and\ \bibinfo
  {author} {\bibfnamefont {J.-G.}\ \bibnamefont {Park}},\ }\href {\doibase
  10.1038/srep20904} {\bibfield  {journal} {\bibinfo  {journal} {Sci. Rep.}\
  }\textbf {\bibinfo {volume} {6}},\ \bibinfo {pages} {20904} (\bibinfo {year}
  {2016})}\BibitemShut {NoStop}%
\bibitem [{\citenamefont {Lee}\ \emph {et~al.}(2016)\citenamefont {Lee},
  \citenamefont {Choi}, \citenamefont {Lee}, \citenamefont {Park},\ and\
  \citenamefont {Park}}]{Lee2016}%
  \BibitemOpen
  \bibfield  {author} {\bibinfo {author} {\bibfnamefont {S.}~\bibnamefont
  {Lee}}, \bibinfo {author} {\bibfnamefont {K.-Y.}\ \bibnamefont {Choi}},
  \bibinfo {author} {\bibfnamefont {S.}~\bibnamefont {Lee}}, \bibinfo {author}
  {\bibfnamefont {B.~H.}\ \bibnamefont {Park}}, \ and\ \bibinfo {author}
  {\bibfnamefont {J.-G.}\ \bibnamefont {Park}},\ }\href {\doibase
  10.1063/1.4961211} {\bibfield  {journal} {\bibinfo  {journal} {APL Mater.}\
  }\textbf {\bibinfo {volume} {4}},\ \bibinfo {pages} {086108} (\bibinfo {year}
  {2016})}\BibitemShut {NoStop}%
\bibitem [{\citenamefont {Joy}\ and\ \citenamefont
  {Vasudevan}(1992)}]{PhysRevB.46.5134}%
  \BibitemOpen
  \bibfield  {author} {\bibinfo {author} {\bibfnamefont {P.~A.}\ \bibnamefont
  {Joy}}\ and\ \bibinfo {author} {\bibfnamefont {S.}~\bibnamefont
  {Vasudevan}},\ }\href {\doibase 10.1103/PhysRevB.46.5134} {\bibfield
  {journal} {\bibinfo  {journal} {Phys. Rev. B}\ }\textbf {\bibinfo {volume}
  {46}},\ \bibinfo {pages} {5134} (\bibinfo {year} {1992})}\BibitemShut
  {NoStop}%
\bibitem [{\citenamefont {Susner}\ \emph {et~al.}(2017)\citenamefont {Susner},
  \citenamefont {Chyasnavichyus}, \citenamefont {McGuire}, \citenamefont
  {Ganesh},\ and\ \citenamefont {Maksymovych}}]{Susner2017}%
  \BibitemOpen
  \bibfield  {author} {\bibinfo {author} {\bibfnamefont {M.~A.}\ \bibnamefont
  {Susner}}, \bibinfo {author} {\bibfnamefont {M.}~\bibnamefont
  {Chyasnavichyus}}, \bibinfo {author} {\bibfnamefont {M.~A.}\ \bibnamefont
  {McGuire}}, \bibinfo {author} {\bibfnamefont {P.}~\bibnamefont {Ganesh}}, \
  and\ \bibinfo {author} {\bibfnamefont {P.}~\bibnamefont {Maksymovych}},\
  }\href {\doibase 10.1002/adma.201602852} {\bibfield  {journal} {\bibinfo
  {journal} {Adv. Mater.}\ }\textbf {\bibinfo {volume} {29}},\ \bibinfo {pages}
  {1602852} (\bibinfo {year} {2017})}\BibitemShut {NoStop}%
\bibitem [{\citenamefont {Ressouche}\ \emph {et~al.}(2010)\citenamefont
  {Ressouche}, \citenamefont {Loire}, \citenamefont {Simonet}, \citenamefont
  {Ballou}, \citenamefont {Stunault},\ and\ \citenamefont
  {Wildes}}]{PhysRevB.82.100408}%
  \BibitemOpen
  \bibfield  {author} {\bibinfo {author} {\bibfnamefont {E.}~\bibnamefont
  {Ressouche}}, \bibinfo {author} {\bibfnamefont {M.}~\bibnamefont {Loire}},
  \bibinfo {author} {\bibfnamefont {V.}~\bibnamefont {Simonet}}, \bibinfo
  {author} {\bibfnamefont {R.}~\bibnamefont {Ballou}}, \bibinfo {author}
  {\bibfnamefont {A.}~\bibnamefont {Stunault}}, \ and\ \bibinfo {author}
  {\bibfnamefont {A.}~\bibnamefont {Wildes}},\ }\href {\doibase
  10.1103/PhysRevB.82.100408} {\bibfield  {journal} {\bibinfo  {journal} {Phys.
  Rev. B}\ }\textbf {\bibinfo {volume} {82}},\ \bibinfo {pages} {100408(R)}
  (\bibinfo {year} {2010})}\BibitemShut {NoStop}%
\bibitem [{\citenamefont {Lan{\c{c}}on}\ \emph {et~al.}(2016)\citenamefont
  {Lan{\c{c}}on}, \citenamefont {Walker}, \citenamefont {Ressouche},
  \citenamefont {Ouladdiaf}, \citenamefont {Rule}, \citenamefont {McIntyre},
  \citenamefont {Hicks}, \citenamefont {R{\o}nnow},\ and\ \citenamefont
  {Wildes}}]{Lancon2016}%
  \BibitemOpen
  \bibfield  {author} {\bibinfo {author} {\bibfnamefont {D.}~\bibnamefont
  {Lan{\c{c}}on}}, \bibinfo {author} {\bibfnamefont {H.~C.}\ \bibnamefont
  {Walker}}, \bibinfo {author} {\bibfnamefont {E.}~\bibnamefont {Ressouche}},
  \bibinfo {author} {\bibfnamefont {B.}~\bibnamefont {Ouladdiaf}}, \bibinfo
  {author} {\bibfnamefont {K.~C.}\ \bibnamefont {Rule}}, \bibinfo {author}
  {\bibfnamefont {G.~J.}\ \bibnamefont {McIntyre}}, \bibinfo {author}
  {\bibfnamefont {T.~J.}\ \bibnamefont {Hicks}}, \bibinfo {author}
  {\bibfnamefont {H.~M.}\ \bibnamefont {R{\o}nnow}}, \ and\ \bibinfo {author}
  {\bibfnamefont {A.~R.}\ \bibnamefont {Wildes}},\ }\href {\doibase
  10.1103/PhysRevB.94.214407} {\bibfield  {journal} {\bibinfo  {journal} {Phys.
  Rev. B}\ }\textbf {\bibinfo {volume} {94}},\ \bibinfo {pages} {214407}
  (\bibinfo {year} {2016})}\BibitemShut {NoStop}%
\bibitem [{\citenamefont {Murayama}\ \emph {et~al.}(2016)\citenamefont
  {Murayama}, \citenamefont {Okabe}, \citenamefont {Urushihara}, \citenamefont
  {Asaka}, \citenamefont {Fukuda}, \citenamefont {Isobe}, \citenamefont
  {Yamamoto},\ and\ \citenamefont {Matsushita}}]{Murayama2016}%
  \BibitemOpen
  \bibfield  {author} {\bibinfo {author} {\bibfnamefont {C.}~\bibnamefont
  {Murayama}}, \bibinfo {author} {\bibfnamefont {M.}~\bibnamefont {Okabe}},
  \bibinfo {author} {\bibfnamefont {D.}~\bibnamefont {Urushihara}}, \bibinfo
  {author} {\bibfnamefont {T.}~\bibnamefont {Asaka}}, \bibinfo {author}
  {\bibfnamefont {K.}~\bibnamefont {Fukuda}}, \bibinfo {author} {\bibfnamefont
  {M.}~\bibnamefont {Isobe}}, \bibinfo {author} {\bibfnamefont
  {K.}~\bibnamefont {Yamamoto}}, \ and\ \bibinfo {author} {\bibfnamefont
  {Y.}~\bibnamefont {Matsushita}},\ }\href
  {http://dx.doi.org/10.1063/1.4961712} {\bibfield  {journal} {\bibinfo
  {journal} {J. Appl. Phys.}\ }\textbf {\bibinfo {volume} {120}},\ \bibinfo
  {pages} {142114} (\bibinfo {year} {2016})}\BibitemShut {NoStop}%
\bibitem [{\citenamefont {Wildes}\ \emph {et~al.}(1998)\citenamefont {Wildes},
  \citenamefont {Roessli}, \citenamefont {Lebech},\ and\ \citenamefont
  {Godfrey}}]{Wildes_1998}%
  \BibitemOpen
  \bibfield  {author} {\bibinfo {author} {\bibfnamefont {A.~R.}\ \bibnamefont
  {Wildes}}, \bibinfo {author} {\bibfnamefont {B.}~\bibnamefont {Roessli}},
  \bibinfo {author} {\bibfnamefont {B.}~\bibnamefont {Lebech}}, \ and\ \bibinfo
  {author} {\bibfnamefont {K.~W.}\ \bibnamefont {Godfrey}},\ }\href {\doibase
  10.1088/0953-8984/10/28/020} {\bibfield  {journal} {\bibinfo  {journal} {J.
  Phys. Condens.}\ }\textbf {\bibinfo {volume} {10}},\ \bibinfo {pages} {6417}
  (\bibinfo {year} {1998})}\BibitemShut {NoStop}%
\bibitem [{\citenamefont {Kurosawa}\ \emph {et~al.}(1983)\citenamefont
  {Kurosawa}, \citenamefont {Saito},\ and\ \citenamefont
  {Yamaguchi}}]{Kurosawa83}%
  \BibitemOpen
  \bibfield  {author} {\bibinfo {author} {\bibfnamefont {K.}~\bibnamefont
  {Kurosawa}}, \bibinfo {author} {\bibfnamefont {S.}~\bibnamefont {Saito}}, \
  and\ \bibinfo {author} {\bibfnamefont {Y.}~\bibnamefont {Yamaguchi}},\
  }\href@noop {} {\bibfield  {journal} {\bibinfo  {journal} {J. Phys. Soc.
  Japan}\ }\textbf {\bibinfo {volume} {52}},\ \bibinfo {pages} {3919} (\bibinfo
  {year} {1983})}\BibitemShut {NoStop}%
\bibitem [{\citenamefont {Masubuchi}\ \emph {et~al.}(2008)\citenamefont
  {Masubuchi}, \citenamefont {Hoya}, \citenamefont {Watanabe}, \citenamefont
  {Takahashi}, \citenamefont {Ban}, \citenamefont {Ohkubo}, \citenamefont
  {Takase},\ and\ \citenamefont {Takano}}]{Masubuchi2008}%
  \BibitemOpen
  \bibfield  {author} {\bibinfo {author} {\bibfnamefont {T.}~\bibnamefont
  {Masubuchi}}, \bibinfo {author} {\bibfnamefont {H.}~\bibnamefont {Hoya}},
  \bibinfo {author} {\bibfnamefont {T.}~\bibnamefont {Watanabe}}, \bibinfo
  {author} {\bibfnamefont {Y.}~\bibnamefont {Takahashi}}, \bibinfo {author}
  {\bibfnamefont {S.}~\bibnamefont {Ban}}, \bibinfo {author} {\bibfnamefont
  {N.}~\bibnamefont {Ohkubo}}, \bibinfo {author} {\bibfnamefont
  {K.}~\bibnamefont {Takase}}, \ and\ \bibinfo {author} {\bibfnamefont
  {Y.}~\bibnamefont {Takano}},\ }\href {\doibase 10.1016/j.jallcom.2007.06.063}
  {\bibfield  {journal} {\bibinfo  {journal} {J. Alloys Compd.}\ }\textbf
  {\bibinfo {volume} {460}},\ \bibinfo {pages} {668} (\bibinfo {year}
  {2008})}\BibitemShut {NoStop}%
\bibitem [{\citenamefont {Manr{\'{i}}quez}\ \emph {et~al.}(2000)\citenamefont
  {Manr{\'{i}}quez}, \citenamefont {Barahona},\ and\ \citenamefont
  {Pe{\~{n}}a}}]{Manriquez2000}%
  \BibitemOpen
  \bibfield  {author} {\bibinfo {author} {\bibfnamefont {V.}~\bibnamefont
  {Manr{\'{i}}quez}}, \bibinfo {author} {\bibfnamefont {P.}~\bibnamefont
  {Barahona}}, \ and\ \bibinfo {author} {\bibfnamefont {O.}~\bibnamefont
  {Pe{\~{n}}a}},\ }\href {\doibase 10.1016/S0025-5408(00)00384-6} {\bibfield
  {journal} {\bibinfo  {journal} {Mater. Res. Bull.}\ }\textbf {\bibinfo
  {volume} {35}},\ \bibinfo {pages} {1889} (\bibinfo {year}
  {2000})}\BibitemShut {NoStop}%
\bibitem [{\citenamefont {He}\ \emph {et~al.}(2003)\citenamefont {He},
  \citenamefont {Dai}, \citenamefont {Huang}, \citenamefont {Lin},\ and\
  \citenamefont {Hsia}}]{He2003}%
  \BibitemOpen
  \bibfield  {author} {\bibinfo {author} {\bibfnamefont {Y.}~\bibnamefont
  {He}}, \bibinfo {author} {\bibfnamefont {Y.~D.}\ \bibnamefont {Dai}},
  \bibinfo {author} {\bibfnamefont {H.}~\bibnamefont {Huang}}, \bibinfo
  {author} {\bibfnamefont {J.}~\bibnamefont {Lin}}, \ and\ \bibinfo {author}
  {\bibfnamefont {Y.}~\bibnamefont {Hsia}},\ }\href {\doibase
  10.1016/S0925-8388(03)00196-8} {\bibfield  {journal} {\bibinfo  {journal} {J.
  Alloys Compd.}\ }\textbf {\bibinfo {volume} {359}},\ \bibinfo {pages} {41}
  (\bibinfo {year} {2003})}\BibitemShut {NoStop}%
\bibitem [{\citenamefont {Takano}\ \emph {et~al.}(2003)\citenamefont {Takano},
  \citenamefont {Arai}, \citenamefont {Takahashi}, \citenamefont {Takase},\
  and\ \citenamefont {Sekizawa}}]{Takano2003}%
  \BibitemOpen
  \bibfield  {author} {\bibinfo {author} {\bibfnamefont {Y.}~\bibnamefont
  {Takano}}, \bibinfo {author} {\bibfnamefont {A.}~\bibnamefont {Arai}},
  \bibinfo {author} {\bibfnamefont {Y.}~\bibnamefont {Takahashi}}, \bibinfo
  {author} {\bibfnamefont {K.}~\bibnamefont {Takase}}, \ and\ \bibinfo {author}
  {\bibfnamefont {K.}~\bibnamefont {Sekizawa}},\ }\href {\doibase
  10.1063/1.1539078} {\bibfield  {journal} {\bibinfo  {journal} {J. Appl.
  Phys.}\ }\textbf {\bibinfo {volume} {93}},\ \bibinfo {pages} {8197} (\bibinfo
  {year} {2003})}\BibitemShut {NoStop}%
\bibitem [{\citenamefont {Bhutani}\ \emph {et~al.}(2020)\citenamefont
  {Bhutani}, \citenamefont {Zuo}, \citenamefont {McAuliffe}, \citenamefont
  {{dela Cruz}},\ and\ \citenamefont {Shoemaker}}]{Bhutani2020}%
  \BibitemOpen
  \bibfield  {author} {\bibinfo {author} {\bibfnamefont {A.}~\bibnamefont
  {Bhutani}}, \bibinfo {author} {\bibfnamefont {J.~L.}\ \bibnamefont {Zuo}},
  \bibinfo {author} {\bibfnamefont {R.~D.}\ \bibnamefont {McAuliffe}}, \bibinfo
  {author} {\bibfnamefont {C.~R.}\ \bibnamefont {{dela Cruz}}}, \ and\ \bibinfo
  {author} {\bibfnamefont {D.~P.}\ \bibnamefont {Shoemaker}},\ }\href {\doibase
  10.1103/PhysRevMaterials.4.034411} {\bibfield  {journal} {\bibinfo  {journal}
  {Phys. Rev. Mater.}\ }\textbf {\bibinfo {volume} {4}},\ \bibinfo {pages}
  {34411} (\bibinfo {year} {2020})}\BibitemShut {NoStop}%
\bibitem [{\citenamefont {Hewat}(1986)}]{D2Bpaper}%
  \BibitemOpen
  \bibfield  {author} {\bibinfo {author} {\bibfnamefont {A.~W.}\ \bibnamefont
  {Hewat}},\ }\href@noop {} {\bibfield  {journal} {\bibinfo  {journal} {Mater.
  Sci. Forum}\ }\textbf {\bibinfo {volume} {9}},\ \bibinfo {pages} {69}
  (\bibinfo {year} {1986})}\BibitemShut {NoStop}%
\bibitem [{\citenamefont {Suard}\ and\ \citenamefont {Wildes}(2019)}]{D2Bdata}%
  \BibitemOpen
  \bibfield  {author} {\bibinfo {author} {\bibfnamefont {E.}~\bibnamefont
  {Suard}}\ and\ \bibinfo {author} {\bibfnamefont {A.~R.}\ \bibnamefont
  {Wildes}},\ }\href@noop {} {}\bibinfo {howpublished} {ILL, Refer to ILL
  experiment number EASY-541} (\bibinfo {year} {2019})\BibitemShut {NoStop}%
\bibitem [{\citenamefont {Toby}\ and\ \citenamefont
  {Von~Dreele}(2013)}]{GSAS2}%
  \BibitemOpen
  \bibfield  {author} {\bibinfo {author} {\bibfnamefont {B.}~\bibnamefont
  {Toby}}\ and\ \bibinfo {author} {\bibfnamefont {R.~B.}\ \bibnamefont
  {Von~Dreele}},\ }\href@noop {} {\bibfield  {journal} {\bibinfo  {journal} {J.
  Appl. Crystallogr.}\ }\textbf {\bibinfo {volume} {46}},\ \bibinfo {pages}
  {544} (\bibinfo {year} {2013})}\BibitemShut {NoStop}%
\bibitem [{\citenamefont {Wildes}\ \emph {et~al.}(2015)\citenamefont {Wildes},
  \citenamefont {Simonet}, \citenamefont {Ressouche}, \citenamefont {McIntyre},
  \citenamefont {Avdeev}, \citenamefont {Suard}, \citenamefont {Kimber},
  \citenamefont {Lan{\c{c}}on}, \citenamefont {Pepe}, \citenamefont
  {Moubaraki},\ and\ \citenamefont {Hicks}}]{Wildes2015}%
  \BibitemOpen
  \bibfield  {author} {\bibinfo {author} {\bibfnamefont {A.~R.}\ \bibnamefont
  {Wildes}}, \bibinfo {author} {\bibfnamefont {V.}~\bibnamefont {Simonet}},
  \bibinfo {author} {\bibfnamefont {E.}~\bibnamefont {Ressouche}}, \bibinfo
  {author} {\bibfnamefont {G.~J.}\ \bibnamefont {McIntyre}}, \bibinfo {author}
  {\bibfnamefont {M.}~\bibnamefont {Avdeev}}, \bibinfo {author} {\bibfnamefont
  {E.}~\bibnamefont {Suard}}, \bibinfo {author} {\bibfnamefont {S.~A.~J.}\
  \bibnamefont {Kimber}}, \bibinfo {author} {\bibfnamefont {D.}~\bibnamefont
  {Lan{\c{c}}on}}, \bibinfo {author} {\bibfnamefont {G.}~\bibnamefont {Pepe}},
  \bibinfo {author} {\bibfnamefont {B.}~\bibnamefont {Moubaraki}}, \ and\
  \bibinfo {author} {\bibfnamefont {T.~J.}\ \bibnamefont {Hicks}},\ }\href
  {\doibase 10.1103/PhysRevB.92.224408} {\bibfield  {journal} {\bibinfo
  {journal} {Phys. Rev. B}\ }\textbf {\bibinfo {volume} {92}},\ \bibinfo
  {pages} {224408} (\bibinfo {year} {2015})}\BibitemShut {NoStop}%
\bibitem [{\citenamefont {Stewart}\ \emph {et~al.}(2009)\citenamefont
  {Stewart}, \citenamefont {Deen}, \citenamefont {Andersen}, \citenamefont
  {Schober}, \citenamefont {Barth{\'{e}}l{\'{e}}my}, \citenamefont {Hillier},
  \citenamefont {Murani}, \citenamefont {Hayes},\ and\ \citenamefont
  {Lindenau}}]{Stewart:db5048}%
  \BibitemOpen
  \bibfield  {author} {\bibinfo {author} {\bibfnamefont {J.~R.}\ \bibnamefont
  {Stewart}}, \bibinfo {author} {\bibfnamefont {P.~P.}\ \bibnamefont {Deen}},
  \bibinfo {author} {\bibfnamefont {K.~H.}\ \bibnamefont {Andersen}}, \bibinfo
  {author} {\bibfnamefont {H.}~\bibnamefont {Schober}}, \bibinfo {author}
  {\bibfnamefont {J.-F.}\ \bibnamefont {Barth{\'{e}}l{\'{e}}my}}, \bibinfo
  {author} {\bibfnamefont {J.~M.}\ \bibnamefont {Hillier}}, \bibinfo {author}
  {\bibfnamefont {A.~P.}\ \bibnamefont {Murani}}, \bibinfo {author}
  {\bibfnamefont {T.}~\bibnamefont {Hayes}}, \ and\ \bibinfo {author}
  {\bibfnamefont {B.}~\bibnamefont {Lindenau}},\ }\href {\doibase
  10.1107/S0021889808039162} {\bibfield  {journal} {\bibinfo  {journal} {J.
  Appl. Crystallogr.}\ }\textbf {\bibinfo {volume} {42}},\ \bibinfo {pages}
  {69} (\bibinfo {year} {2009})}\BibitemShut {NoStop}%
\bibitem [{\citenamefont {Wildes}\ \emph {et~al.}(2019)\citenamefont {Wildes},
  \citenamefont {Coak}, \citenamefont {Graham},\ and\ \citenamefont
  {Saxena}}]{D7data}%
  \BibitemOpen
  \bibfield  {author} {\bibinfo {author} {\bibfnamefont {A.~R.}\ \bibnamefont
  {Wildes}}, \bibinfo {author} {\bibfnamefont {M.~J.}\ \bibnamefont {Coak}},
  \bibinfo {author} {\bibfnamefont {J.~N.}\ \bibnamefont {Graham}}, \ and\
  \bibinfo {author} {\bibfnamefont {S.~S.}\ \bibnamefont {Saxena}},\
  }\href@noop {} {}\bibinfo {howpublished} {ILL,
  https://doi.ill.fr/10.5291/ILL-DATA.5-32-870} (\bibinfo {year}
  {2019})\BibitemShut {NoStop}%
\bibitem [{\citenamefont {Schweika}(2010)}]{Schweika_2010}%
  \BibitemOpen
  \bibfield  {author} {\bibinfo {author} {\bibfnamefont {W.}~\bibnamefont
  {Schweika}},\ }\href {\doibase 10.1088/1742-6596/211/1/012026} {\bibfield
  {journal} {\bibinfo  {journal} {Journal of Physics: Conference Series}\
  }\textbf {\bibinfo {volume} {211}},\ \bibinfo {pages} {012026} (\bibinfo
  {year} {2010})}\BibitemShut {NoStop}%
\bibitem [{\citenamefont {Rodriguez-Carvajal}(1993)}]{Fullprof}%
  \BibitemOpen
  \bibfield  {author} {\bibinfo {author} {\bibfnamefont {J.}~\bibnamefont
  {Rodriguez-Carvajal}},\ }\href@noop {} {\bibfield  {journal} {\bibinfo
  {journal} {Physica B.}\ }\textbf {\bibinfo {volume} {192}} (\bibinfo {year}
  {1993})}\BibitemShut {NoStop}%
\bibitem [{\citenamefont {Paddison}\ and\ \citenamefont
  {Goodwin}(2012)}]{PhysRevLett.108.017204}%
  \BibitemOpen
  \bibfield  {author} {\bibinfo {author} {\bibfnamefont {J.~A.~M.}\
  \bibnamefont {Paddison}}\ and\ \bibinfo {author} {\bibfnamefont {A.~L.}\
  \bibnamefont {Goodwin}},\ }\href {\doibase 10.1103/PhysRevLett.108.017204}
  {\bibfield  {journal} {\bibinfo  {journal} {Phys. Rev. Lett.}\ }\textbf
  {\bibinfo {volume} {108}},\ \bibinfo {pages} {017204} (\bibinfo {year}
  {2012})}\BibitemShut {NoStop}%
\bibitem [{\citenamefont {Paddison}\ \emph {et~al.}(2013)\citenamefont
  {Paddison}, \citenamefont {Stewart},\ and\ \citenamefont
  {Goodwin}}]{Paddison_2013}%
  \BibitemOpen
  \bibfield  {author} {\bibinfo {author} {\bibfnamefont {J.~A.~M.}\
  \bibnamefont {Paddison}}, \bibinfo {author} {\bibfnamefont {J.~R.}\
  \bibnamefont {Stewart}}, \ and\ \bibinfo {author} {\bibfnamefont {A.~L.}\
  \bibnamefont {Goodwin}},\ }\href {\doibase 10.1088/0953-8984/25/45/454220}
  {\bibfield  {journal} {\bibinfo  {journal} {J. Phys. Condens.}\ }\textbf
  {\bibinfo {volume} {25}},\ \bibinfo {pages} {454220} (\bibinfo {year}
  {2013})}\BibitemShut {NoStop}%
\bibitem [{\citenamefont {Paddison}(2019)}]{PaddisonScatty}%
  \BibitemOpen
  \bibfield  {author} {\bibinfo {author} {\bibfnamefont {J.~A.~M.}\
  \bibnamefont {Paddison}},\ }\href {\doibase 10.1107/S2053273318015632}
  {\bibfield  {journal} {\bibinfo  {journal} {Acta Cryst. A}\ }\textbf
  {\bibinfo {volume} {75}},\ \bibinfo {pages} {14} (\bibinfo {year}
  {2019})}\BibitemShut {NoStop}%
\bibitem [{\citenamefont {Bain}\ and\ \citenamefont {Berry}(2008)}]{Bain2008}%
  \BibitemOpen
  \bibfield  {author} {\bibinfo {author} {\bibfnamefont {G.~A.}\ \bibnamefont
  {Bain}}\ and\ \bibinfo {author} {\bibfnamefont {J.~F.}\ \bibnamefont
  {Berry}},\ }\href {\doibase 10.1021/ed085p532} {\bibfield  {journal}
  {\bibinfo  {journal} {J. Chem. Educ.}\ }\textbf {\bibinfo {volume} {85}},\
  \bibinfo {pages} {532} (\bibinfo {year} {2008})}\BibitemShut {NoStop}%
\bibitem [{\citenamefont {Brown}(1979)}]{Magff}%
  \BibitemOpen
  \bibfield  {author} {\bibinfo {author} {\bibfnamefont {P.~J.}\ \bibnamefont
  {Brown}},\ }\href@noop {} {\emph {\bibinfo {title} {Electron and
  Magnetisation Densities in Molecules and Crystals}}}\ (\bibinfo  {publisher}
  {Plenum Press N. Y.},\ \bibinfo {year} {1979})\ Chap.\ \bibinfo {chapter}
  {Magnetic Neutron Scattering}\BibitemShut {NoStop}%
\bibitem [{\citenamefont {Wildes}\ \emph {et~al.}(2012)\citenamefont {Wildes},
  \citenamefont {Rule}, \citenamefont {Bewley}, \citenamefont {Enderle},\ and\
  \citenamefont {Hicks}}]{Wildes12}%
  \BibitemOpen
  \bibfield  {author} {\bibinfo {author} {\bibfnamefont {A.~R.}\ \bibnamefont
  {Wildes}}, \bibinfo {author} {\bibfnamefont {K.~C.}\ \bibnamefont {Rule}},
  \bibinfo {author} {\bibfnamefont {R.~I.}\ \bibnamefont {Bewley}}, \bibinfo
  {author} {\bibfnamefont {M.}~\bibnamefont {Enderle}}, \ and\ \bibinfo
  {author} {\bibfnamefont {T.~J.}\ \bibnamefont {Hicks}},\ }\href@noop {}
  {\bibfield  {journal} {\bibinfo  {journal} {J. Phys.: Condens. Matter}\
  }\textbf {\bibinfo {volume} {24}},\ \bibinfo {pages} {416004} (\bibinfo
  {year} {2012})}\BibitemShut {NoStop}%
\bibitem [{\citenamefont {Goossens}\ \emph {et~al.}(2000)\citenamefont
  {Goossens}, \citenamefont {Studer}, \citenamefont {Kennedy},\ and\
  \citenamefont {Hicks}}]{Goossens2000}%
  \BibitemOpen
  \bibfield  {author} {\bibinfo {author} {\bibfnamefont {D.~J.}\ \bibnamefont
  {Goossens}}, \bibinfo {author} {\bibfnamefont {A.~J.}\ \bibnamefont
  {Studer}}, \bibinfo {author} {\bibfnamefont {S.~J.}\ \bibnamefont {Kennedy}},
  \ and\ \bibinfo {author} {\bibfnamefont {T.~J.}\ \bibnamefont {Hicks}},\
  }\href {\doibase 10.1088/0953-8984/12/18/308} {\bibfield  {journal} {\bibinfo
   {journal} {J. Phys. Condens. Matter}\ }\textbf {\bibinfo {volume} {12}},\
  \bibinfo {pages} {4233} (\bibinfo {year} {2000})}\BibitemShut {NoStop}%
\bibitem [{\citenamefont {Rule}\ \emph {et~al.}(2002)\citenamefont {Rule},
  \citenamefont {Kennedy}, \citenamefont {Goossens}, \citenamefont {Mulders},\
  and\ \citenamefont {Hicks}}]{Rule2002}%
  \BibitemOpen
  \bibfield  {author} {\bibinfo {author} {\bibfnamefont {K.~C.}\ \bibnamefont
  {Rule}}, \bibinfo {author} {\bibfnamefont {S.~J.}\ \bibnamefont {Kennedy}},
  \bibinfo {author} {\bibfnamefont {D.~J.}\ \bibnamefont {Goossens}}, \bibinfo
  {author} {\bibfnamefont {A.~M.}\ \bibnamefont {Mulders}}, \ and\ \bibinfo
  {author} {\bibfnamefont {T.~J.}\ \bibnamefont {Hicks}},\ }\href {\doibase
  10.1007/s003390201363} {\bibfield  {journal} {\bibinfo  {journal} {Appl.
  Phys. A}\ }\textbf {\bibinfo {volume} {74}},\ \bibinfo {pages} {s811}
  (\bibinfo {year} {2002})}\BibitemShut {NoStop}%
\bibitem [{\citenamefont {Rule}\ \emph {et~al.}(2003)\citenamefont {Rule},
  \citenamefont {Ersez}, \citenamefont {Kennedy},\ and\ \citenamefont
  {Hicks}}]{Rule2003}%
  \BibitemOpen
  \bibfield  {author} {\bibinfo {author} {\bibfnamefont {K.~C.}\ \bibnamefont
  {Rule}}, \bibinfo {author} {\bibfnamefont {T.}~\bibnamefont {Ersez}},
  \bibinfo {author} {\bibfnamefont {S.~J.}\ \bibnamefont {Kennedy}}, \ and\
  \bibinfo {author} {\bibfnamefont {T.~J.}\ \bibnamefont {Hicks}},\ }\href
  {\doibase 10.1016/S0921-4526(03)00179-0} {\bibfield  {journal} {\bibinfo
  {journal} {Physica B Condens. Matter}\ }\textbf {\bibinfo {volume} {335}},\
  \bibinfo {pages} {6} (\bibinfo {year} {2003})}\BibitemShut {NoStop}%
\bibitem [{\citenamefont {Wildes}\ \emph {et~al.}(1994)\citenamefont {Wildes},
  \citenamefont {Kennedy},\ and\ \citenamefont {Hicks}}]{Wildes1994}%
  \BibitemOpen
  \bibfield  {author} {\bibinfo {author} {\bibfnamefont {A.~R.}\ \bibnamefont
  {Wildes}}, \bibinfo {author} {\bibfnamefont {S.~J.}\ \bibnamefont {Kennedy}},
  \ and\ \bibinfo {author} {\bibfnamefont {T.~J.}\ \bibnamefont {Hicks}},\
  }\href@noop {} {\bibfield  {journal} {\bibinfo  {journal} {J. Phys. Condens.
  Matter}\ }\textbf {\bibinfo {volume} {6}},\ \bibinfo {pages} {L335} (\bibinfo
  {year} {1994})}\BibitemShut {NoStop}%
\end{thebibliography}%

\end{document}